\documentclass{article}
\usepackage[utf8]{inputenc}

\usepackage{geometry}
\usepackage[super]{natbib}
\setcitestyle{comma} 

\usepackage[hidelinks]{hyperref}
\hypersetup{
    colorlinks=true,
    linkcolor=blue, 
    filecolor=black,
    citecolor=red,
    urlcolor=blue
    }

\usepackage{graphicx}

\usepackage{amssymb}
\usepackage{amsmath}

\usepackage{enumitem}
\setlist{
    nosep, 
    after=\vspace{.5em}, 
    before=\vspace{.5em}
}

\usepackage{lineno}

\usepackage{authblk}

\usepackage{changepage} 
  {\end{adjustwidth}}

\title{Modeling the amplification of epidemic spread by individuals exposed to misinformation on social media}

\author[1,$\dagger$]{Matthew R. DeVerna}
\author[2]{Francesco Pierri}
\author[1]{Yong-Yeol Ahn}
\author[1]{Santo Fortunato}
\author[1]{Alessandro Flammini}
\author[1]{Filippo Menczer}

\affil[1]{\small\textit{Luddy School of Informatics, Computing, and Engineering, Indiana University, Bloomington, USA}}
\affil[2]{\small \textit{Department of Electronics, Information and Bioengineering, Politecnico di Milano, Milano, Italy}}
\affil[$\dagger$]{\small \textit{Corresponding author. Email:} \texttt{mdeverna@iu.edu}}

\date{
\small This preprint has not yet undergone peer review.\\
\today
}

\begin{document}

\maketitle

\begin{abstract}
    Understanding how misinformation affects the spread of disease is crucial for public health, especially given recent research indicating that misinformation can increase vaccine hesitancy and discourage vaccine uptake.
    However, it is difficult to investigate the interaction between misinformation and epidemic outcomes due to the dearth of data-informed holistic epidemic models. 
    Here, we employ an epidemic model that incorporates a large, mobility-informed physical contact network as well as the distribution of misinformed individuals across counties derived from social media data. 
    The model allows us to simulate various scenarios to understand how epidemic spreading can be affected by misinformation spreading through one particular social media platform. 
    Using this model, we compare a worst-case scenario, in which individuals become misinformed after a single exposure to low-credibility content, to a best-case scenario where the population is highly resilient to misinformation. 
    We estimate the additional portion of the U.S. population that would become infected over the course of the COVID-19 epidemic in the worst-case scenario. 
    This work can provide policymakers with insights about the potential harms of exposure to online vaccine misinformation.  
\end{abstract}

\bigskip
\bigskip
\bigskip

\clearpage

Social factors, such as information sharing, play a crucial role in shaping the dynamics and epidemiology of infectious diseases\cite{Buckee2021Jul, Bavel2020May}.
For instance, a population's willingness to adopt public health measures (or lack thereof) largely determines their successes or failures\cite{Mitze2020Dec, Bauch2013Oct}. 
A population's behavioral response to outbreaks can be influenced by mass media, as witnessed during the 2009 H1N1 influenza pandemic\cite{Pol11}, or by social media and the anti-vaccination movement\cite{Allen2023Oct, Gallotti2020, Broniatowski2018, burki2019vaccine}.

\begin{figure}
    \centerline{
    \includegraphics[width=0.75\textwidth]{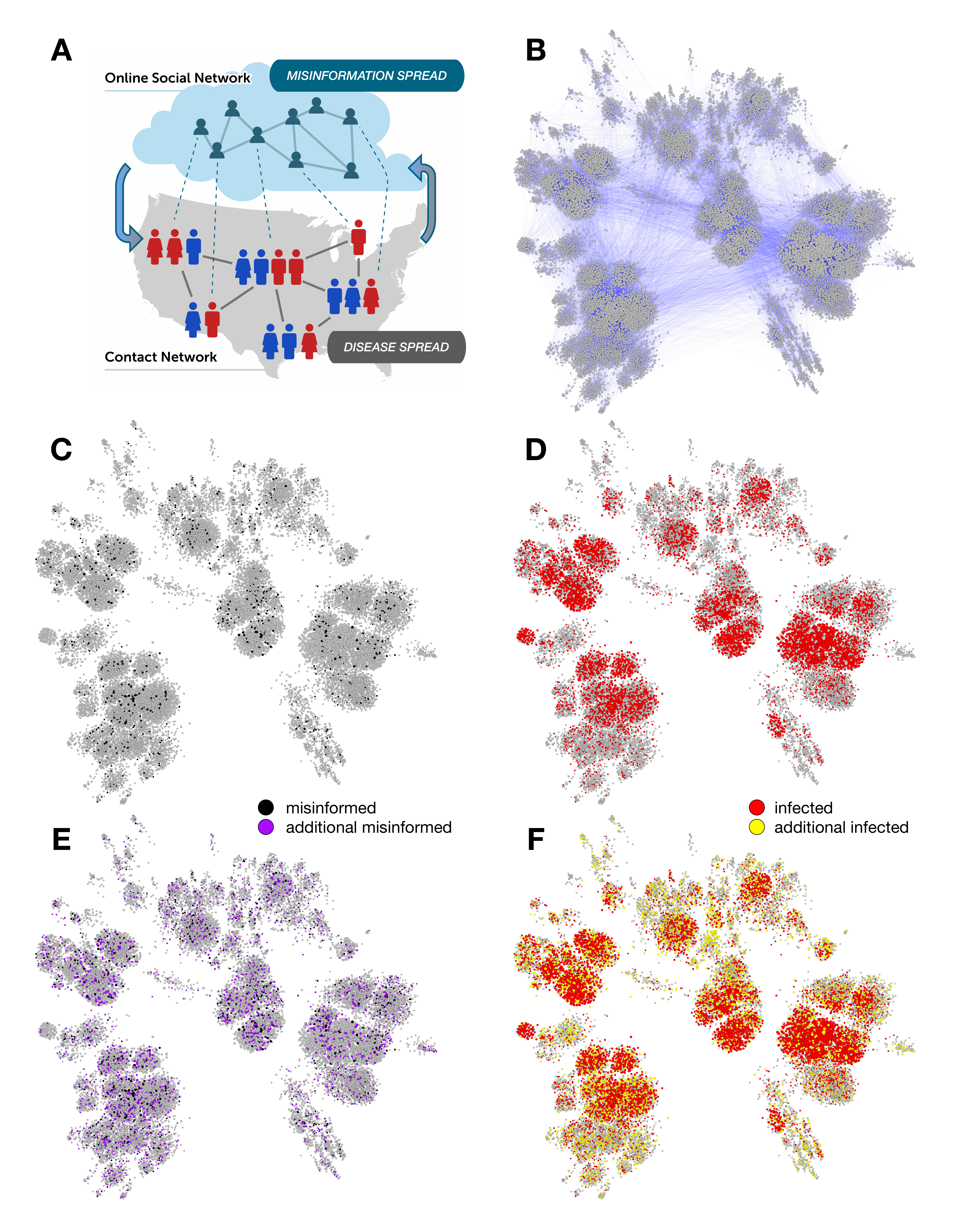}
    }
    \caption{\small The spread of misinformation affects the transmission of disease. 
    (\textbf{A})~Schematic illustration of the misinformation and contact networks.
    Online social networks foster misinformation dissemination while physical contact networks, such as those that connect co-workers in an office or pupils in a school, facilitate disease transmission.
    Dotted links indicate that the same people participate in both networks, which have different topologies; e.g., the information network tends to have stronger political homophily while the contact network tends to have stronger geographic homophily. 
    We focus on the impact of misinformation spread on disease transmission (downward arrow), while the opposite effect (upward arrow, e.g., individuals ceasing to share misinformation due to illness) falls outside the scope of this investigation.
    (\textbf{B})~A contact network based on 0.01\% county population samples. Nodes are sized based on degree (number of contacts). In a scenario with limited spread of misinformation (black nodes in \textbf{C}), the simulations of disease spread leads to a number of infected individuals (red nodes in \textbf{D}). In a scenario where the misinformation spreads more widely (purple nodes in \textbf{E}), more individuals get infected (yellow nodes in \textbf{F}).
    }
    \label{fig:2nets}
\end{figure}

A great deal of work has explored how to model the influence of human behavior on the spread of infectious diseases\cite{Verelst2016Dec, Funk2010Sep}. 
Here we focus on risky behaviors affecting disease transmission that are associated with misinformed individuals. 
Misinformation spreading on social networks has been linked to poor compliance with COVID-19 public health guidance\cite{Roozenbeek2020Oct}.
Greater exposure to unreliable news articles about COVID-19 vaccines has been linked to an increase in vaccine hesitancy and a decrease in vaccination rates at both state and county levels in the United States\cite{rathje2022social, Pierri2022Apr}.
Exposure to online misinformation has also been shown to increase vaccine hesitancy in laboratory experiments\cite{loomba2021misinfo}.
This is particularly detrimental during vaccination campaigns as clusters of individuals adopting anti-vaccination opinions can make it challenging for a population to reach herd immunity\cite{Chan2021May, Salathe2011Oct}.
Proper management of epidemic crises in the modern age thus requires the understanding of the complex relationship between the spread of (mis)information through online social networks and the spread of disease through physical contact networks (Fig.~\ref{fig:2nets}).

Agent-based simulations have shown that misinformation may impede the suppression of epidemics in various ways\cite{Sontag2022Mar, Mumtaz2022Apr, Prandi2020Dec, Brainard2019Nov}. 
One model estimated that between March and November 2021, misinformation caused at least 198 thousand additional COVID-19 cases, 2,800 additional deaths, and \$299M in additional hospital costs in Canada\cite{CCA2023FaultLines}. 
However, there is a growing need to strengthen the connections between simulation results and real-world outcomes by integrating real-world data from social media\cite{Bedson2021Jul, Sooknanan2020Jul}. 

We address this challenge by proposing a data-informed epidemic model that takes both the distribution of misinformed individuals and a physical mobility network into account. 
Using this data, we augment the Susceptible Infected Recovered (SIR) model to account for a subpopulation of ``misinformed'' individuals. 
We refer to this as the Susceptible Misinformed Infected Recovered (SMIR) model.
We explore how the misinformed group can affect the larger, ordinary population using a multi-level agent-based simulation based on two large, data-informed networks: a social network where misinformation spreads and a contact network where the disease can propagate.
A contact network of approximately 20 million nodes is constructed by leveraging large-scale Twitter data, county-level voting records, and cell phone mobility data.
We incorporate theoretically extreme values of the parameter responsible for the epidemic transmission to evaluate best- and worst-case scenarios about the impact of misinformed individuals on the spread of disease and obtain quantitative bounds on the harm caused by misinformation. 
The proposed model lets us move beyond simplified experimental settings to assess the impacts of misinformation\cite{Tay2024Jan}. 

\section*{Results}
\label{sec:results}

\begin{figure}
    \centerline{\includegraphics[width=\textwidth]{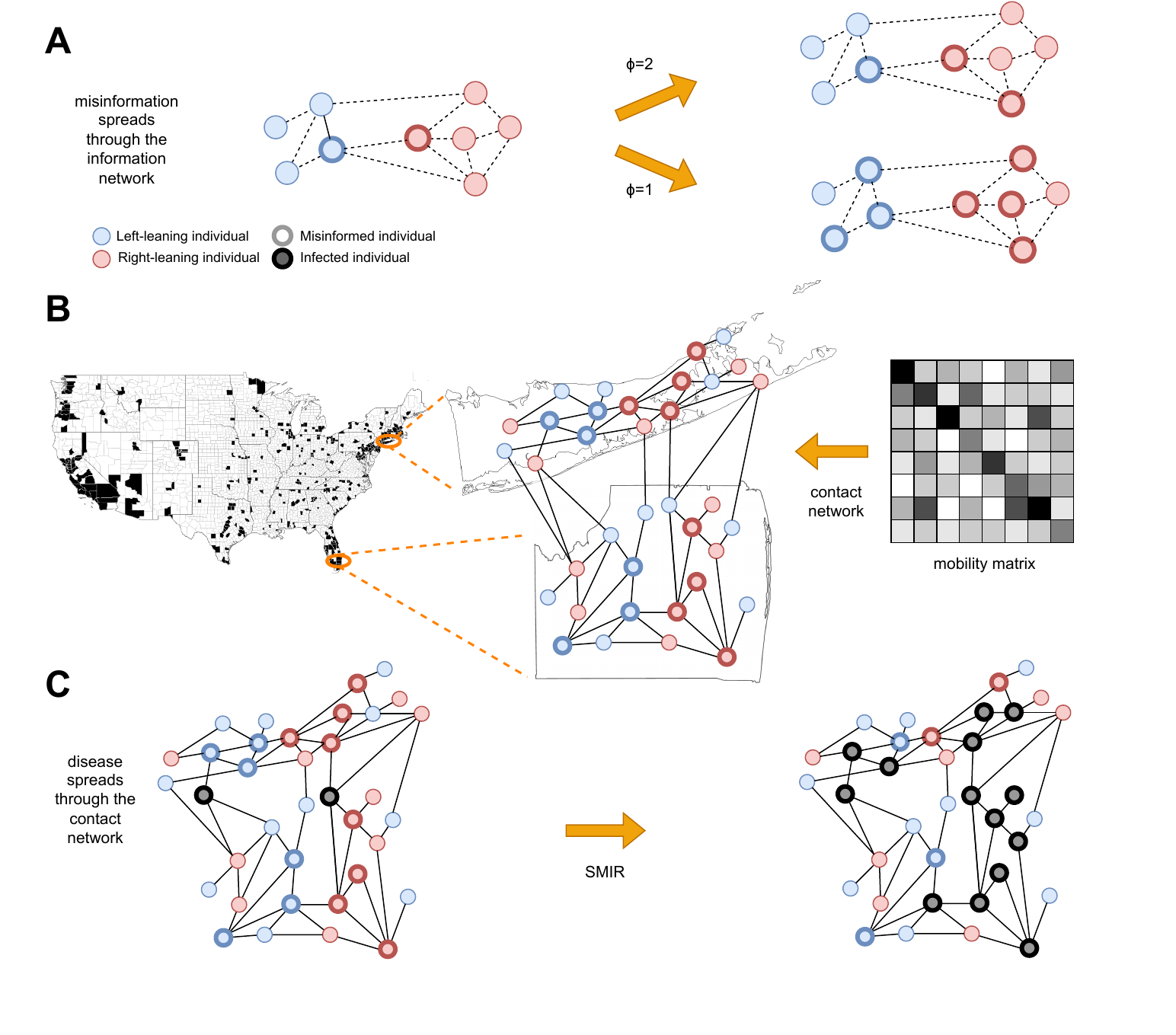}}
    \caption{\small An idealized example of our multi-level modeling framework. (\textbf{A}) Spread of misinformation through an information network (dashed lines). Colors represent ideological homophily. Nodes with bold borders are misinformed about the epidemic. The misinformation spreads through a complex contagion (linear threshold) model; two scenarios show that a lower threshold $\phi$ leads to more misinformed nodes. (\textbf{B}) Construction of the contact network (solid lines) for counties with sufficient information diffusion data (in black) to provide reasonable estimates about the fraction of misinformed individuals. Note that these counties account for 63.52\% of U.S. voters. Each location's population size and ideological mix are based on empirical data, and misinformed individuals are based on the information diffusion model. Links among individuals within and between locations are based on empirical mobility data. (\textbf{C}) The infection spreads through the contact network (black nodes), according to the SMIR model.}
    \label{fig:model}
\end{figure}

We utilize a multi-level, agent-based model to examine the influence of misinformation on epidemic spread. 
Our approach combines an empirically derived information network with a contact network calibrated with real-world data, as illustrated in Fig.~\ref{fig:model}.
Information diffusion is modeled by leveraging a large set of users of a popular social media platform. Epidemic simulations are subsequently conducted on contact networks populated with misinformed individuals.

We start from a large collection of English-language discussions taking place on Twitter about COVID-19 vaccines\cite{deverna2021covaxxy}.
From approximately nine months of this data (Jan. 4--Sep. 30, 2021), we geolocate over 2 million U.S. users who shared almost 26 million tweets and focus on accounts in 341 U.S. counties containing more than 200 Twitter users each. 
We also infer an account's political alignment and whether they shared any likely misinformation (see Methods). 
Twitter is not representative of the U.S. population, and people also access information in other ways, such as traditional media and word of mouth. 
However, this social media platform serves as one large, realistic network through which people share information about the disease. 

With this data, we build a directed and weighted information diffusion network, in which an edge $(i \rightarrow j, w)$ indicates that $j$ retweeted $i$ $w$ times. 
There are various ways to model the infodemic\cite{DAndrea2022Epidemic}.
We simulate the spread of misinformation on this network, as illustrated in Fig.~\ref{fig:model}A.
Accounts that share or reshare posts containing misinformation are considered misinformed.
These accounts serve as the initial \emph{seeds} from which misinformation proliferates, with exposure to this content likely concentrated within the wider network\cite{Budak2024Jun}.
Many users may not actively participate in content sharing; for instance, only about half of U.S. Twitter users engage in sharing\cite{Pew2022Mar}.
Even without active sharing, exposure to misinformation or misleading content can still influence individual behavior\cite{loomba2021misinfo, Allen2023Oct}. 

To account for users who may be misinformed through exposure, we employ a single-step linear threshold opinion-spreading process\cite{Granovetter1978May}.
While many social influence models have been proposed\cite{Fortunato2009review}, this is a simple way to capture complex contagion, according to which individuals may require multiple exposures to misinformation before they become misinformed themselves\cite{centola2010spread,Lilian2013srep,monsted2017evidence}. 
Let a linear threshold $\phi$ represent the minimum number of misinformed friends needed for an ordinary node to become misinformed.
If the total number of misinformed friends of $i$ is greater than or equal to $\phi$, $i$ is marked as misinformed ($M$). 
The remaining nodes are marked as ordinary susceptibles ($O$). 
We can interpret $\phi$ as a measure of ``resilience'' to misinformation; as it increases, individuals require more exposure to misinformation to be converted to the misinformed group. 
Conversely, we can think of $\phi$ as inversely related to intent or motivation to engage with low-credibility content\cite{Simon2021Jul}. 
Note that since we explore the full range of $\phi$ values, the following results are unaffected whether the threshold is defined based on the number of users or the number of retweets. 

\begin{figure}
    \includegraphics[width=\linewidth]{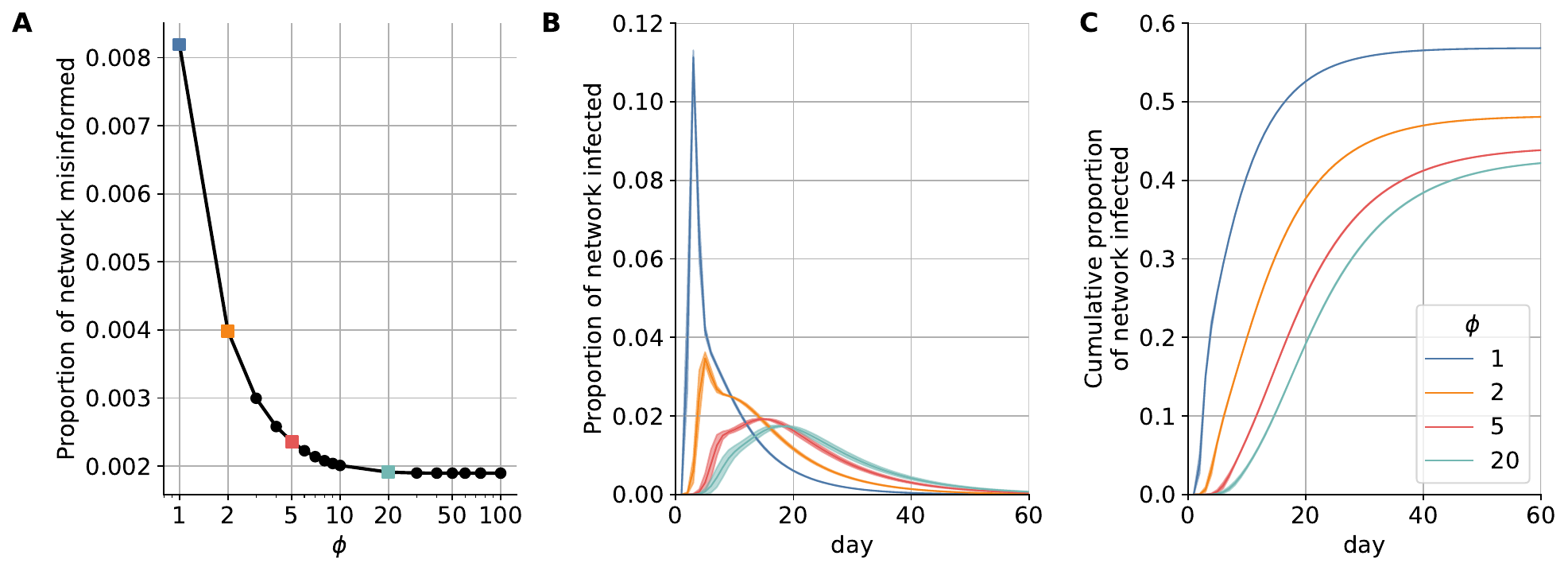} 
    \caption{\small More misinformed individuals lead to a larger portion of the network becoming infected. 
    Decreasing the resilience $\phi$ (\textbf{A}) increases the size of the misinformed subpopulation, leading to (\textbf{B}) faster infection spreading and (\textbf{C}) a greater cumulative number of infections. 
    In panels (\textbf{B}, \textbf{C}), lines and corresponding shaded regions represent the mean and standard deviation across simulations, respectively.}
  \label{fig:effect_of_phi}
\end{figure}

Fig.~\ref{fig:effect_of_phi}A shows how $\phi$ influences the number of misinformed individuals within the retweet network.
With strong resilience ($\phi > 10$), exposure to misinformation does not have much effect and few nodes are converted to the misinformed group.
Conversely, when resilience to misinformation is very low (as in the simple contagion case $\phi = 1$), all nodes exposed to a misinformation-containing post are converted to the misinformed group.
Through this process, empirically observed misinformation-sharing behavior leads to information networks with misinformed subpopulations of varying sizes based on different $\phi$ values.

We generate contact networks for different thresholds ($1 \leq \phi \leq 20$) to compare the impact of misinformed subpopulations of different sizes.
Given a threshold $\phi$ and the corresponding information network, we aim to construct a physical contact network containing empirically calibrated misinformed subpopulations (Fig.~\ref{fig:model}B). 
The process begins by selecting a sample of individuals from each county within the information network. 
As party affiliation has been identified as a risk factor associated with excess mortality during the COVID-19 pandemic\cite{JacobWallace2023Jul}, county samples are constructed to match the percentage of Republicans and Democrats who voted in the 2020 U.S. presidential election. 
For each county, we add the sampled nodes to the physical network marked as misinformed ($M$) or ordinary susceptible ($O$), based on their label within the retweet network.
Sampling with replacement allows us to select individuals such that the overall proportions of Republicans and Democrats match the voting records. 
A 10\% sample leads to $N \approx 20$ million nodes. 
A network based on a much smaller sample is illustrated in Fig.~\ref{fig:2nets}B. 
This process captures empirical measurements of the ideological split, relative population size, and quantity of misinformed individuals in each county. 
It also allows us to account for the known link between the ideological motivations of users and their exposure to misinformation\cite{Budak2024Jun, Pierri2022Apr}. 
We add contact network edges by leveraging cell phone mobility data that provides the probability of an individual traveling within and between counties.
See Methods for details.

Disease-spreading dynamics on the contact network are simulated using the SMIR model (Fig.~\ref{fig:model}C). 
As in the standard SIR\cite{anderson1991infectious}, a parameter $\beta$ describes the average number of infected individuals generated by an infected individual in a time unit.
We can express $\beta = p \bar{k}$ in terms of two critical parameters that affect the spreading dynamics: the density of the contact network, captured by its average degree $\bar{k}$, and the transmission probability $p$.   
Infected individuals recover with rate $\gamma$.

We extend this epidemic model to account for misinformed and ordinary subpopulations. 
Ordinary individuals are considered to be well-informed about public health guidelines, such as social distancing, mask-wearing, and vaccination.
Mitigation measures such as social distancing decrease $\bar{k}$, while those such as masking and vaccination decrease $p$. 
Misinformed individuals, having been exposed to untrustworthy information, are assumed to be less likely to follow these recommended behaviors, thereby increasing the risk of infection for themselves and others\cite{DAndrea2022Feb}. 
A simple way to model the combined effects of misinformation on these behaviors through a single parameter is to set $\bar{k}=25$, a high value corresponding to the average number of daily contacts prior to the COVID-19 pandemic\cite{LiuSocialContactPatterns}, and use extreme values of $p$ to capture worst- and best-case scenarios. 
An effective reduction of contacts, resulting for example from social distancing or lockdowns, can be captured by decreasing the $p$ parameter. 

We therefore model the refusal of any mitigation measures by selecting the maximum value $p_M = 1$ for misinformed individuals. 
In contrast, we model the adoption of several mitigation measures by selecting an extremely small value $p_O = 0.01$ for ordinary individuals. 
The former scenario portrays a realistic number of interactions during non-pandemic times, accompanied by high transmission rates due to the absence of preventive measures, such as social distancing, mask-wearing, or vaccinations.
The latter demonstrates decreased daily interactions and reduced transmission rates resulting from the implementation of these preventive measures.
Using the empirically calibrated contact networks in conjunction with these extreme parameters, the simulation approach allows us to bound the best- and worst-case scenarios in a data-informed manner (see Methods for more information). 

The effects of the misinformed subpopulation size on the daily incidence of infection (illustrated in Fig.~\ref{fig:2nets}C-F on a small network) are quantified in Fig.~\ref{fig:effect_of_phi}B on a large network (10\% sample). 
The worst case capturing a heavily misinformed population ($\phi=1$) corresponds to an additional 9\% of the population being infected at peak time (a six-fold increase) compared to a resilient population following expert guidance in the best-case scenario ($\phi=20$).
The peak also occurs approximately two weeks earlier. 
The cumulative effect is also significant, with an additional 14\% of the population infected over the course of the epidemic compared to case with a more resilient population --- a 32\% relative increase (Fig.~\ref{fig:effect_of_phi}C). 

We explored alternative scenarios for the ratio $p_M/p_O$ through a mean-field approximation.
Predictably, as this ratio gets larger, the infected population increases and the peak infection occurs earlier. 
We also considered different sample sizes for the empirical network and found that the main results are robust. 
These analyses can be found in Supplementary Information.

\section*{Discussion}
\label{sec:Discussion}

Exposure to online health misinformation is associated with risky behaviors such as vaccine hesitancy and refusal\cite{Pierri2022Apr}.
There is also experimental evidence suggesting a causal link\cite{loomba2021misinfo, Thaker2021, Allen2023Oct}.
While one study found no evidence that misinformation reduces intent to vaccinate, the authors report that they did not have sufficient power to detect small effects\cite{Porter2023Mar}. 
Assuming an association exists between exposure to health misinformation on one particular social media platform and risky behaviors, this work uses large-scale epidemic simulations to further link the behaviors of misinformed individuals to an accelerated spread of disease. 
Our model is anchored in empirical data\cite{Bedson2021Jul, Sooknanan2020Jul} to explore potential outcomes.

Agent-based simulations of the SMIR model let us study the epidemic on empirically calibrated contact networks. 
By comparing a worst-case scenario, in which individuals become misinformed after a single exposure to low-credibility content, to a best-case scenario where the population is highly resilient to misinformation, the model estimates that the peak of the infection is amplified by a factor of six and accelerated by two weeks.
This would result in an additional 14\% of the population becoming infected --- nearly 47 million Americans based on recent U.S. Census data\cite{UScensusPopClock}. 
The corresponding price tag of vaccine misinformation would be over \$143B, using estimated health care costs associated with COVID-19 in the U.S.\cite{doi:10.1377/hlthaff.2020.00426}.

While these figures are based on extreme scenarios, they represent an alarming bound on the harm of exposure to online vaccine misinformation. 
They should provide public health authorities as well as social media platforms with heightened motivation to curb vaccine misinformation, despite the difficulties posed by social media design\cite{doi:10.1126/sciadv.adh2132}.

Our results do not address the differential effects of the epidemic on the two populations of ordinary and misinformed individuals.
We carry out such an analysis using a mean-field approximation of the model, which assumes all individuals have an equal chance of interacting. 
The mean-field model demonstrates how the risky behaviors of misinformed individuals can adversely impact those following public health guidelines, worsening outcomes for the entire population (see Supplementary Information). 
Additionally, we use the mean-field model to explore the role of homophily in the population, i.e., scenarios where misinformed individuals are more likely to be connected to other misinformed individuals and similarly for the ordinary population.
We find that increasing homophily can benefit the overall population by protecting ordinary citizens; however, it may also lead to higher infection rates within the misinformed subpopulation (see Supplementary Information). 

We acknowledge several limitations in our approach. 
The model assumes the existence of a causal link between exposure to online misinformation and the adoption of risky behaviors.
There is a need for models that can provide support for this assumption beyond existing lab experiments\cite{loomba2021misinfo, Allen2023Oct}. 

Using empirical retweet data as a proxy for social connections may not capture potential passive exposure to misinformation.
While follower relationships could diminish this limitation, our choice allows us to focus on users who are more likely to be impacted due to their active engagement. 

We model a single wave of infection with somewhat arbitrary extreme-case parameters ($p_O = 0.01, p_M = 1$). A broader range of values is explored in a mean-field scenario, along with the effect of the size of the misinformed population (see Supplementary Information). 
Of course, as $p_O \rightarrow 0$, only the misinformed population can get infected in the model. 
However, since the mean-field scenario ignores the network structure, its results cannot be directly compared to those of the agent-based model. 
COVID-19 saw multiple waves of infection with different variants, varying reproduction numbers, levels of immunity, and so on. 
Future work should attempt to quantify the potential effects of misinformation in more realistic scenarios, where the key parameters $p_M$ and $p_O$ could be calibrated on empirical surveillance data from particular regions and time periods.  

We also assume uniform resilience to misinformation for all individuals during the information diffusion process, although this attribute likely differs across individuals. 
Future directions could involve more sophisticated models to account for these heterogeneities. 
For instance, cognitive models of misinformation acceptance\cite{Borukhson2022Jun} could be incorporated into the simulation with misinformation exposure data collected from social media.
Such integration would enable the transition of individuals from ordinary to misinformed susceptible states throughout the simulation, allowing for a simultaneous examination of opinion and disease dynamics. 
Some theoretical models have already explored similar approaches and obtained results that align with our findings\cite{Mumtaz2022Apr, Sontag2022Mar}.

Finally, although individual beliefs and behaviors may vary over time, our model simplifies the scenario by dichotomizing individuals into misinformed and ordinary subpopulations and assuming constant transmission rates. 
Future extensions of the model could account for a feedback loop whereby witnessing local infections could drive changes in behaviors equivalent to the transition of individuals out of the misinformed population\cite{Zeng24QMF}. 

\section*{Methods}
\label{sec:methods}

\paragraph{Twitter and derived data.}
\label{sec:methods:tweetdata}

Twitter posts in the CoVaxxy dataset\cite{deverna2021covaxxy} were collected in real time via the stream/filter endpoint of the Twitter Application Programming Interface (API).
To capture the online discourse surrounding COVID-19 vaccines in English, a comprehensive set of English-language keywords was carefully curated.
Beginning with the initial seeds of ``covid'' and ``vaccine,'' a snowball sampling technique\cite{di2022vaccineu} was used to identify co-occurring relevant keywords in December 2020\cite{deverna2021covaxxy}.
The resulting list contained almost 80 keywords, available online\cite{covaxxy_dataset}.
To confirm the relevance of the collected tweets to the topic of vaccines, we examined the coverage obtained by incrementally adding keywords, starting with the most common ones. 
Over 90\% of the tweets in 2021 contained at least one of the three most common keywords: ``vaccine,'' ``vaccination,'' or ``vaccinate.'' 
To infer the location of accounts, we used the Carmen Python library\cite{dredze2013carmen} that leverages self-reported location metadata within user profiles (embedded in tweets).
As an account's location may change over time (captured across multiple tweets), we utilize the most recent location.
We geolocate 2,047,800 users residing in all 50 U.S. states, who shared a total of 25,806,856 tweets by mapping self-reported locations to U.S. counties.
The information network is constructed from accounts in 341 counties that contain more than 200 Twitter users each.
Political alignment is estimated using a third-party list of annotated news sources\cite{Robertson2023, Robertson2018Nov}.
It is averaged across all the sources shared by each account.
Nodes with an estimated alignment greater (smaller) than zero are considered Republican (Democrat). 
We infer the political alignment of some additional accounts, who did not share links to news sources, using a label-propagation algorithm\cite{Conover2010predicting} on the retweet network.
If all of a node's neighbors have political alignment scores, its score is estimated using the weighted average of its neighbors, with weights based on retweets.
The process is iterated until each node without a score has at least one neighbor without a score. 
Misinformation is defined at the source level.
Tweets containing links to articles from a list of low-credibility sources compiled by NewsGuard (score below 60) are labeled as spreading misinformation. This approach is common practice and has been validated in the literature\cite{Pierri2022Apr, grinberg2019fake, bovet2019influence, lazer2018science, shao2018spread}. 

\paragraph{Contact network edges.}
\label{sec:methods:contact_net}

To construct edges in our contact network, we utilize SafeGraph cell-phone mobility data\cite{yuan2023implications}, which contains information on the number of people residing in over 200K Census-Block-Groups (CBG) who visited 4.3M Points-of-Interest (POI) in the United States.
This data has been widely employed to study human mobility patterns during the COVID pandemic.
We used the average daily number of individuals moving during 2019, as a reference for business-as-usual mobility, and aggregated all CBGs and POIs at the county level. 
This aggregation results in a county-by-county matrix $L$, where each element $L_{xy}$ represents the average daily number of individuals in county $x$ moving to county $y$ or vice versa. 
We then normalized $L_{xy}$ to obtain the average probability of individuals in counties $x$ and $y$ coming into contact, and multiplied by the total number of edges to obtain the expected number of connections between individuals in counties $x$ and $y$:
$E_{xy} = \frac{L_{xy}}{\sum_{x',y'} L_{x'y'}} \frac{\bar{k} N}{2}$
where the sum is over all county pairs and $\frac{\bar{k} N}{2}$ is the total number of edges. 
Next, we create a physical contact network with $N$ nodes by following a procedure akin to a stochastic block model\cite{karrer2011stochastic} used to generate networks with localized communities. 
For each pair of distinct locations $x$ and $y$, we draw $E_{xy}$ edges between random pairs of nodes in $x$ and $y$.   
Additionally, we draw $E_{xx}$ edges among random pairs of individuals within the same location $x$, representing homogeneous mixing within each county.
At the end of the process, the network has the target average degree $\bar{k}$. We use $\bar{k} = 25$ and show how this parameter affects the infections in Supplementary Information.

\paragraph{Simulation details.}
\label{sec:methods:simulation_details}

Agent-based SMIR simulations are initiated by randomly selecting 100 misinformed nodes and designating them as infected. 
The disease spreading dynamics are then simulated for 100 steps, which correspond to days.
To align with COVID-19 dynamics, we utilize the CDC's recommended quarantine period of 5 days as our recovery period\cite{cdc_isolation_2023} ($\gamma = 0.2$).
Each simulation is repeated ten times, and the average outcome is reported.

\paragraph{Data and code availability.}
Code and data are available in a public repository (\href{https://github.com/osome-iu/bounding-misinfo-impact-on-disease-spread}{github.com/osome-iu/bounding-misinfo-impact-on-disease-spread}).

\section*{Acknowledgments}

We are grateful to Yuan Yuan, Marco Ajelli, Alessio Brina, and Brea Perry for helpful discussions.
This work was supported in part by the Swiss National Science Foundation (grant 209250), the National Science Foundation (grants 1927425 and 1927418), the Army Research Office (contract W911NF-21-1-0194), the European Union (NextGenerationEU project PNRR-PE-AI FAIR), the Italian Ministry of Education (PRIN PNRR grant CODE prot. P2022AKRZ9 and PRIN grant DEMON prot. 2022BAXSPY), Knight Foundation, and Craig Newmark Philanthropies.

\section*{Author contributions} 
The initial concept for this research was developed by YA, SF, AF, FM.
The final study design was developed collectively by all authors.
Data collection was led by MRD as part of the CoVaxxy project~\cite{covaxxy_dataset}.
All mean-field modeling was conducted by MRD.
FP developed the early code for the agent-based simulation, which was subsequently refined and expanded by MRD for publication.
Analysis was conducted primarily by MRD, with help from FP, and guidance from YA, SF, AF, FM.
The SMIR model was developed collaboratively by all authors.
Visualizations were created by MRD, FP, and FM with input from all authors.
The first draft was written by MRD, with revisions from all authors.
FM oversaw the progression of the study.

\section*{Competing interests} 
Authors declare that they have no competing interests.

\section*{Ethics}

This study, focusing on public data, poses minimal risk to human subjects. 
Consequently, the Indiana University Institutional Review Board has exempted it from review (protocol number 1102004860).
All data collection and analysis adhered to Twitter's terms of service.

\bibliography{ref.bib}

\begin{thebibliography}{62}
\providecommand{\natexlab}[1]{#1}
\providecommand{\url}[1]{\texttt{#1}}
\expandafter\ifx\csname urlstyle\endcsname\relax
  \providecommand{\doi}[1]{doi: #1}\else
  \providecommand{\doi}{doi: \begingroup \urlstyle{rm}\Url}\fi

\bibitem[Allen et~al.(2023)Allen, Watts, and Rand]{Allen2023Oct}
J.~N.~L. Allen, D.~J. Watts, and D.~Rand.
\newblock {Quantifying the Impact of Misinformation and Vaccine-Skeptical
  Content on Facebook}.
\newblock \emph{PsyArXiv}, Oct. 2023.
\newblock URL \url{https://doi.org/10.31234/osf.io/nwsqa}.

\bibitem[Anderson and May(1991)]{anderson1991infectious}
R.~M. Anderson and R.~M. May.
\newblock \emph{Infectious diseases of humans: dynamics and control}.
\newblock Oxford university press, 1991.

\bibitem[Bartsch et~al.(2020)Bartsch, Ferguson, McKinnell, O'Shea, Wedlock,
  Siegmund, and Lee]{doi:10.1377/hlthaff.2020.00426}
S.~M. Bartsch, M.~C. Ferguson, J.~A. McKinnell, K.~J. O'Shea, P.~T. Wedlock,
  S.~S. Siegmund, and B.~Y. Lee.
\newblock {The Potential Health Care Costs And Resource Use Associated With
  COVID-19 In The United States}.
\newblock \emph{Health Affairs}, 39\penalty0 (6):\penalty0 927--935, 2020.
\newblock \doi{10.1377/hlthaff.2020.00426}.
\newblock URL \url{https://doi.org/10.1377/hlthaff.2020.00426}.

\bibitem[Bauch and Galvani(2013)]{Bauch2013Oct}
C.~T. Bauch and A.~P. Galvani.
\newblock {Social Factors in Epidemiology}.
\newblock \emph{Science}, 342\penalty0 (6154):\penalty0 47--49, Oct. 2013.
\newblock URL \url{https://doi.org/10.1126/science.1244492}.

\bibitem[Bavel et~al.(2020)Bavel, Baicker, Boggio, Capraro, Cichocka, Cikara,
  Crockett, Crum, Douglas, Druckman, Drury, Dube, Ellemers, Finkel, Fowler,
  Gelfand, Han, Haslam, Jetten, Kitayama, Mobbs, Napper, Packer, Pennycook,
  Peters, Petty, Rand, Reicher, Schnall, Shariff, Skitka, Smith, Sunstein,
  Tabri, Tucker, Linden, Lange, Weeden, Wohl, Zaki, Zion, and
  Willer]{Bavel2020May}
J.~J.~V. Bavel, K.~Baicker, P.~S. Boggio, V.~Capraro, A.~Cichocka, M.~Cikara,
  M.~J. Crockett, A.~J. Crum, K.~M. Douglas, J.~N. Druckman, J.~Drury, O.~Dube,
  N.~Ellemers, E.~J. Finkel, J.~H. Fowler, M.~Gelfand, S.~Han, S.~A. Haslam,
  J.~Jetten, S.~Kitayama, D.~Mobbs, L.~E. Napper, D.~J. Packer, G.~Pennycook,
  E.~Peters, R.~E. Petty, D.~G. Rand, S.~D. Reicher, S.~Schnall, A.~Shariff,
  L.~J. Skitka, S.~S. Smith, C.~R. Sunstein, N.~Tabri, J.~A. Tucker, S.~v.~d.
  Linden, P.~v. Lange, K.~A. Weeden, M.~J.~A. Wohl, J.~Zaki, S.~R. Zion, and
  R.~Willer.
\newblock {Using social and behavioural science to support COVID-19 pandemic
  response}.
\newblock \emph{Nat Hum Behav}, 4:\penalty0 460--471, May 2020.
\newblock URL \url{https://doi.org/10.1038/s41562-020-0884-z}.

\bibitem[Bedson et~al.(2021)Bedson, Skrip, Pedi, Abramowitz, Carter, Jalloh,
  Funk, Gobat, Giles-Vernick, Chowell, de~Almeida, Elessawi, Scarpino, Hammond,
  Briand, Epstein, H{\ifmmode\acute{e}\else\'{e}\fi}bert-Dufresne, and
  Althouse]{Bedson2021Jul}
J.~Bedson, L.~A. Skrip, D.~Pedi, S.~Abramowitz, S.~Carter, M.~F. Jalloh,
  S.~Funk, N.~Gobat, T.~Giles-Vernick, G.~Chowell, J.~R. de~Almeida,
  R.~Elessawi, S.~V. Scarpino, R.~A. Hammond, S.~Briand, J.~M. Epstein,
  L.~H{\ifmmode\acute{e}\else\'{e}\fi}bert-Dufresne, and B.~M. Althouse.
\newblock {A review and agenda for integrated disease models including social
  and behavioural factors}.
\newblock \emph{Nat Hum Behav}, 5:\penalty0 834--846, July 2021.
\newblock URL \url{https://doi.org/10.1038/s41562-021-01136-2}.

\bibitem[Borukhson et~al.(2022)Borukhson, Lorenz-Spreen, and
  Ragni]{Borukhson2022Jun}
D.~Borukhson, P.~Lorenz-Spreen, and M.~Ragni.
\newblock {When Does an Individual Accept Misinformation? An Extended
  Investigation Through Cognitive Modeling}.
\newblock \emph{Comput Brain Behav}, 5\penalty0 (2):\penalty0 244--260, June
  2022.
\newblock URL \url{https://doi.org/10.1007/s42113-022-00136-3}.

\bibitem[Bovet and Makse(2019)]{bovet2019influence}
A.~Bovet and H.~A. Makse.
\newblock {Influence of fake news in Twitter during the 2016 US presidential
  election}.
\newblock \emph{Nature communications}, 10\penalty0 (1):\penalty0 7, 2019.
\newblock URL \url{https://doi.org/10.1038/s41467-018-07761-2}.

\bibitem[Brainard and Hunter(2019)]{Brainard2019Nov}
J.~Brainard and P.~R. Hunter.
\newblock {Misinformation making a disease outbreak worse: outcomes compared
  for influenza, monkeypox, and norovirus}.
\newblock \emph{SIMULATION}, 96\penalty0 (4):\penalty0 365--374, Nov. 2019.
\newblock URL \url{https:/doi.org/10.1177/0037549719885021}.

\bibitem[Broniatowski et~al.(2018)Broniatowski, Jamison, Qi, AlKulaib, Chen,
  Benton, Quinn, and Dredze]{Broniatowski2018}
D.~A. Broniatowski, A.~M. Jamison, S.~Qi, L.~AlKulaib, T.~Chen, A.~Benton,
  S.~C. Quinn, and M.~Dredze.
\newblock Weaponized health communication: Twitter bots and russian trolls
  amplify the vaccine debate.
\newblock \emph{American Journal of Public Health}, 108\penalty0 (10):\penalty0
  1378--1384, 2018.
\newblock \doi{10.2105/AJPH.2018.304567}.
\newblock URL \url{https://doi.org/10.2105/AJPH.2018.304567}.

\bibitem[Broniatowski et~al.(2023)Broniatowski, Simons, Gu, Jamison, and
  Abroms]{doi:10.1126/sciadv.adh2132}
D.~A. Broniatowski, J.~R. Simons, J.~Gu, A.~M. Jamison, and L.~C. Abroms.
\newblock {The efficacy of Facebook’s vaccine misinformation policies and
  architecture during the COVID-19 pandemic}.
\newblock \emph{Science Advances}, 9\penalty0 (37):\penalty0 eadh2132, 2023.
\newblock \doi{10.1126/sciadv.adh2132}.
\newblock URL \url{https://www.science.org/doi/abs/10.1126/sciadv.adh2132}.

\bibitem[Buckee et~al.(2021)Buckee, Noor, and Sattenspiel]{Buckee2021Jul}
C.~Buckee, A.~Noor, and L.~Sattenspiel.
\newblock {Thinking clearly about social aspects of infectious disease
  transmission}.
\newblock \emph{Nature}, 595:\penalty0 205--213, July 2021.
\newblock URL \url{https://doi.org/10.1038/s41586-021-03694-x}.

\bibitem[Budak et~al.(2024)Budak, Nyhan, Rothschild, Thorson, and
  Watts]{Budak2024Jun}
C.~Budak, B.~Nyhan, D.~M. Rothschild, E.~Thorson, and D.~J. Watts.
\newblock {Misunderstanding the harms of online misinformation}.
\newblock \emph{Nature}, 630:\penalty0 45--53, June 2024.
\newblock \doi{10.1038/s41586-024-07417-w}.
\newblock URL \url{https://doi.org/10.1038/s41586-024-07417-w}.

\bibitem[Burki(2019)]{burki2019vaccine}
T.~Burki.
\newblock Vaccine misinformation and social media.
\newblock \emph{The Lancet Digital Health}, 1\penalty0 (6):\penalty0
  e258--e259, 2019.
\newblock URL \url{https://doi.org/10.1016/S2589-7500(19)30136-0}.

\bibitem[Castellano et~al.(2009)Castellano, Fortunato, and
  Loreto]{Fortunato2009review}
C.~Castellano, S.~Fortunato, and V.~Loreto.
\newblock Statistical physics of social dynamics.
\newblock \emph{Rev. Mod. Phys.}, 81:\penalty0 591--646, 2009.
\newblock \doi{10.1103/RevModPhys.81.591}.
\newblock URL \url{https://link.aps.org/doi/10.1103/RevModPhys.81.591}.

\bibitem[{Centers for Disease Control and
  Prevention}(2023)]{cdc_isolation_2023}
{Centers for Disease Control and Prevention}.
\newblock {Isolation and Precautions for People with COVID-19}, 2023.
\newblock URL
  \url{https://www.cdc.gov/coronavirus/2019-ncov/your-health/isolation.html}.
\newblock Accessed 2023-06-27.

\bibitem[Centola(2010)]{centola2010spread}
D.~Centola.
\newblock The spread of behavior in an online social network experiment.
\newblock \emph{Science}, 329\penalty0 (5996):\penalty0 1194--1197, 2010.

\bibitem[Chan et~al.(2021)Chan, Chiu, Zuo, Wang, Liu, and Hong]{Chan2021May}
H.-W. Chan, C.~P.-Y. Chiu, S.~Zuo, X.~Wang, L.~Liu, and Y.-y. Hong.
\newblock {Not-so-straightforward links between believing in COVID-19-related
  conspiracy theories and engaging in disease-preventive behaviours}.
\newblock \emph{Humanit Soc Sci Commun}, 8\penalty0 (104):\penalty0 1--10, May
  2021.
\newblock URL \url{https://doi.org/10.1057/s41599-021-00781-2}.

\bibitem[Conover et~al.(2011)Conover, Gon\c{c}alves, Ratkiewicz, Flammini, and
  Menczer]{Conover2010predicting}
M.~Conover, B.~Gon\c{c}alves, J.~Ratkiewicz, A.~Flammini, and F.~Menczer.
\newblock Predicting the political alignment of twitter users.
\newblock In \emph{Proceedings of 3rd IEEE Conference on Social Computing
  (SocialCom)}, pages 192--199, 2011.
\newblock \doi{10.1109/PASSAT/SocialCom.2011.34}.

\bibitem[D{'}Andrea et~al.(2022{\natexlab{a}})D{'}Andrea, Artime, Castaldo,
  Sacco, Gallotti, and De~Domenico]{DAndrea2022Epidemic}
V.~D{'}Andrea, O.~Artime, N.~Castaldo, P.~Sacco, R.~Gallotti, and
  M.~De~Domenico.
\newblock {Epidemic proximity and imitation dynamics drive infodemic waves
  during the COVID-19 pandemic}.
\newblock \emph{Phys Rev Res}, 4\penalty0 (1):\penalty0 013158, Feb.
  2022{\natexlab{a}}.
\newblock \doi{10.1103/PhysRevResearch.4.013158}.
\newblock URL \url{https://doi.org/10.1103/PhysRevResearch.4.013158}.

\bibitem[D{'}Andrea et~al.(2022{\natexlab{b}})D{'}Andrea, Gallotti, Castaldo,
  and De~Domenico]{DAndrea2022Feb}
V.~D{'}Andrea, R.~Gallotti, N.~Castaldo, and M.~De~Domenico.
\newblock {Individual risk perception and empirical social structures shape the
  dynamics of infectious disease outbreaks}.
\newblock \emph{PLoS Comput Biol}, 18\penalty0 (2):\penalty0 e1009760, Feb.
  2022{\natexlab{b}}.
\newblock \doi{10.1371/journal.pcbi.1009760}.
\newblock URL \url{https://doi.org/10.1371/journal.pcbi.1009760}.

\bibitem[DeVerna et~al.(2021{\natexlab{a}})DeVerna, Pierri, Truong,
  Bollenbacher, Axelrod, Loynes, Torres-Lugo, Yang, Menczer, and
  Bryden]{covaxxy_dataset}
M.~R. DeVerna, F.~Pierri, B.~T. Truong, J.~Bollenbacher, D.~Axelrod, N.~Loynes,
  C.~Torres-Lugo, K.-C. Yang, F.~Menczer, and J.~Bryden.
\newblock {CoVaxxy Tweet IDs dataset}.
\newblock Zenodo, Feb. 2021{\natexlab{a}}.
\newblock URL \url{https://doi.org/10.5281/zenodo.7752586}.

\bibitem[DeVerna et~al.(2021{\natexlab{b}})DeVerna, Pierri, Truong,
  Bollenbacher, Axelrod, Loynes, Torres-Lugo, Yang, Menczer, and
  Bryden]{deverna2021covaxxy}
M.~R. DeVerna, F.~Pierri, B.~T. Truong, J.~Bollenbacher, D.~Axelrod, N.~Loynes,
  C.~Torres-Lugo, K.-C. Yang, F.~Menczer, and J.~Bryden.
\newblock {CoVaxxy: A Collection of English-Language Twitter Posts About
  COVID-19 Vaccines}.
\newblock In \emph{Proceedings of the International AAAI Conference on Web and
  Social Media}, volume~15, pages 992--999, 2021{\natexlab{b}}.
\newblock URL \url{https://doi.org/10.1609/icwsm.v15i1.18122}.

\bibitem[Di~Giovanni et~al.(2022)Di~Giovanni, Pierri, Torres-Lugo, and
  Brambilla]{di2022vaccineu}
M.~Di~Giovanni, F.~Pierri, C.~Torres-Lugo, and M.~Brambilla.
\newblock {VaccinEU: COVID-19 vaccine conversations on Twitter in French,
  German and Italian}.
\newblock In \emph{Proceedings of the International AAAI Conference on Web and
  Social Media}, volume~16, pages 1236--1244, 2022.
\newblock URL \url{https://doi.org/10.1609/icwsm.v16i1.19374}.

\bibitem[Dredze et~al.(2013)Dredze, Paul, Bergsma, and Tran]{dredze2013carmen}
M.~Dredze, M.~J. Paul, S.~Bergsma, and H.~Tran.
\newblock Carmen: A twitter geolocation system with applications to public
  health.
\newblock In \emph{Proc. AAAI Workshop on Expanding the Boundaries of Health
  Informatics Using AI (HIAI)}, volume~23, page~45, 2013.

\bibitem[Funk et~al.(2010)Funk, Salath\'{e}, and Jansen]{Funk2010Sep}
S.~Funk, M.~Salath\'{e}, and V.~A.~A. Jansen.
\newblock {Modelling the influence of human behaviour on the spread of
  infectious diseases: a review}.
\newblock \emph{J R Soc Interface}, 7\penalty0 (50):\penalty0 1247--1256, Sept.
  2010.
\newblock \doi{10.1098/rsif.2010.0142}.
\newblock URL \url{https://doi.org/10.1098/rsif.2010.0142}.

\bibitem[Gallotti et~al.(2020)Gallotti, Valle, Castaldo, Sacco, and
  De~Domenico]{Gallotti2020}
R.~Gallotti, F.~Valle, N.~Castaldo, P.~Sacco, and M.~De~Domenico.
\newblock {Assessing the risks of `infodemics' in response to COVID-19
  epidemics}.
\newblock \emph{Nature Human Behaviour}, 4\penalty0 (12):\penalty0 1285--1293,
  2020.
\newblock \doi{10.1038/s41562-020-00994-6}.
\newblock URL \url{https://doi.org/10.1038/s41562-020-00994-6}.

\bibitem[Granovetter(1978)]{Granovetter1978May}
M.~Granovetter.
\newblock {Threshold Models of Collective Behavior}.
\newblock \emph{American Journal of Sociology}, May 1978.
\newblock URL \url{https://doi.org/10.1086/226707}.

\bibitem[Grinberg et~al.(2019)Grinberg, Joseph, Friedland, Swire-Thompson, and
  Lazer]{grinberg2019fake}
N.~Grinberg, K.~Joseph, L.~Friedland, B.~Swire-Thompson, and D.~Lazer.
\newblock {Fake news on Twitter during the 2016 US presidential election}.
\newblock \emph{Science}, 363\penalty0 (6425):\penalty0 374--378, 2019.
\newblock URL \url{https://doi.org/10.1126/science.aau2706}.

\bibitem[Himelfarb et~al.(2023)Himelfarb, Boecker, Ève Carignan, Caulfield,
  Cliche, Hodson, Horn, Khenti, Lewandowsky, MacDonald, Mai, Ozawa, and
  Sterling]{CCA2023FaultLines}
A.~Himelfarb, A.~Boecker, M.~Ève Carignan, T.~Caulfield, J.-F. Cliche,
  J.~Hodson, O.~Horn, A.~Khenti, S.~Lewandowsky, N.~MacDonald, P.~Mai,
  S.~Ozawa, and J.~Sterling.
\newblock {Fault Lines: Expert Panel on the Socioeconomic Impacts of Science
  and Health Misinformation}.
\newblock Technical report, Council of Canadian Academies, 2023.
\newblock URL
  \url{https://cca-reports.ca/reports/the-socioeconomic-impacts-of-health-and-science-misinformation}.
\newblock [Online; accessed 20. Jan. 2025].

\bibitem[Jacob~Wallace(2023)]{JacobWallace2023Jul}
P.~Jacob~Wallace.
\newblock {Excess Death Rates for Republican and Democratic Registered Voters
  in Florida and Ohio During the COVID-19}.
\newblock \emph{JAMA Intern Med}, July 2023.
\newblock URL
  \url{https://jamanetwork.com/article.aspx?doi=10.1001/jamainternmed.2023.1154}.

\bibitem[Karrer and Newman(2011)]{karrer2011stochastic}
B.~Karrer and M.~E. Newman.
\newblock Stochastic blockmodels and community structure in networks.
\newblock \emph{Physical Review E}, 83\penalty0 (1):\penalty0 016107, 2011.
\newblock URL \url{https://doi.org/10.1103/PhysRevE.83.016107}.

\bibitem[Lazer et~al.(2018)Lazer, Baum, Benkler, Berinsky, Greenhill, Menczer,
  Metzger, Nyhan, Pennycook, Rothschild, et~al.]{lazer2018science}
D.~M. Lazer, M.~A. Baum, Y.~Benkler, A.~J. Berinsky, K.~M. Greenhill,
  F.~Menczer, M.~J. Metzger, B.~Nyhan, G.~Pennycook, D.~Rothschild, et~al.
\newblock The science of fake news.
\newblock \emph{Science}, 359\penalty0 (6380):\penalty0 1094--1096, 2018.
\newblock URL \url{https://doi.org/10.1126/science.aao2998}.

\bibitem[Liu et~al.(2021)Liu, Berlin, Kiti, Del~Fava, Grow, Zagheni, Melegaro,
  Jenness, Omer, Lopman, and Nelson]{LiuSocialContactPatterns}
C.~Y. Liu, J.~Berlin, M.~C. Kiti, E.~Del~Fava, A.~Grow, E.~Zagheni,
  A.~Melegaro, S.~M. Jenness, S.~B. Omer, B.~Lopman, and K.~Nelson.
\newblock {Rapid Review of Social Contact Patterns During the COVID-19
  Pandemic}.
\newblock \emph{Epidemiology}, 32\penalty0 (6), 2021.
\newblock \doi{10.1097/EDE.0000000000001412}.
\newblock URL \url{https://doi.org/10.1097/EDE.0000000000001412}.

\bibitem[Loomba et~al.(2021)Loomba, de~Figueiredo, Piatek, de~Graaf, and
  Larson]{loomba2021misinfo}
S.~Loomba, A.~de~Figueiredo, S.~J. Piatek, K.~de~Graaf, and H.~J. Larson.
\newblock Measuring the impact of {COVID-19} vaccine misinformation on
  vaccination intent in the {UK} and {USA}.
\newblock \emph{Nature Human Behavior}, 2021.
\newblock URL \url{https://doi.org/10.1038/s41562-021-01056-1}.

\bibitem[Mitze et~al.(2020)Mitze, Kosfeld, Rode, and
  W{\ifmmode\ddot{a}\else\"{a}\fi}lde]{Mitze2020Dec}
T.~Mitze, R.~Kosfeld, J.~Rode, and K.~W{\ifmmode\ddot{a}\else\"{a}\fi}lde.
\newblock {Face masks considerably reduce COVID-19 cases in Germany}.
\newblock \emph{Proc Natl Acad Sci U.S.A}, 117\penalty0 (51):\penalty0
  32293--32301, Dec. 2020.
\newblock \doi{10.1073/pnas.2015954117}.
\newblock URL \url{https://doi.org/10.1073/pnas.2015954117}.

\bibitem[M{\o}nsted et~al.(2017)M{\o}nsted, Sapie{\.z}y{\'n}ski, Ferrara, and
  Lehmann]{monsted2017evidence}
B.~M{\o}nsted, P.~Sapie{\.z}y{\'n}ski, E.~Ferrara, and S.~Lehmann.
\newblock {Evidence of complex contagion of information in social media: An
  experiment using Twitter bots}.
\newblock \emph{PLoS ONE}, 12\penalty0 (9):\penalty0 e0184148, 2017.

\bibitem[Mumtaz et~al.(2022)Mumtaz, Green, and Duggan]{Mumtaz2022Apr}
N.~Mumtaz, C.~Green, and J.~Duggan.
\newblock {Exploring the Effect of Misinformation on Infectious Disease
  Transmission}.
\newblock \emph{Systems}, 10\penalty0 (2):\penalty0 50, Apr. 2022.
\newblock URL \url{https:/doi.org/10.3390/systems10020050}.

\bibitem[Nematzadeh et~al.(2014)Nematzadeh, Ferrara, Flammini, and
  Ahn]{PhysRevLett.113.088701}
A.~Nematzadeh, E.~Ferrara, A.~Flammini, and Y.-Y. Ahn.
\newblock Optimal network modularity for information diffusion.
\newblock \emph{Phys. Rev. Lett.}, 113:\penalty0 088701, 2014.
\newblock \doi{10.1103/PhysRevLett.113.088701}.
\newblock URL \url{https://doi.org/10.1103/PhysRevLett.113.088701}.

\bibitem[Odabaş(2022)]{Pew2022Mar}
M.~Odabaş.
\newblock {5 facts about Twitter `lurkers'}.
\newblock Pew Research Center, 2022.
\newblock URL
  \url{https://www.pewresearch.org/fact-tank/2022/03/16/5-facts-about-twitter-lurkers}.

\bibitem[Pierri et~al.(2022)Pierri, Perry, DeVerna, Yang, Flammini, Menczer,
  and Bryden]{Pierri2022Apr}
F.~Pierri, B.~L. Perry, M.~R. DeVerna, K.-C. Yang, A.~Flammini, F.~Menczer, and
  J.~Bryden.
\newblock {Online misinformation is linked to early COVID-19 vaccination
  hesitancy and refusal}.
\newblock \emph{Sci Rep}, 12\penalty0 (5966):\penalty0 1--7, 2022.
\newblock URL \url{https://doi.org/10.1038/s41598-022-10070-w}.

\bibitem[Poletti et~al.(2011)Poletti, Ajelli, and Merler]{Pol11}
P.~Poletti, M.~Ajelli, and S.~Merler.
\newblock {The effect of risk perception on the 2009 H1N1 pandemic influenza
  dynamics}.
\newblock \emph{PLOS One}, 6\penalty0 (2):\penalty0 e16460, 2011.
\newblock URL \url{https://doi.org/10.1371/journal.pone.0016460}.

\bibitem[Porter et~al.(2023)Porter, Velez, and Wood]{Porter2023Mar}
E.~Porter, Y.~Velez, and T.~J. Wood.
\newblock {Correcting COVID-19 vaccine misinformation in 10 countries}.
\newblock \emph{R. Soc. Open Sci.}, 10\penalty0 (3), Mar. 2023.
\newblock \doi{10.1098/rsos.221097}.

\bibitem[Prandi and Primiero(2020)]{Prandi2020Dec}
L.~Prandi and G.~Primiero.
\newblock {Effects of misinformation diffusion during a pandemic}.
\newblock \emph{Appl Network Sci}, 5\penalty0 (1):\penalty0 1--20, Dec. 2020.
\newblock URL \url{https:/doi.org/10.1007/s41109-020-00327-6}.

\bibitem[Rathje et~al.(2022)Rathje, He, Roozenbeek, Van~Bavel, and van~der
  Linden]{rathje2022social}
S.~Rathje, J.~K. He, J.~Roozenbeek, J.~J. Van~Bavel, and S.~van~der Linden.
\newblock Social media behavior is associated with vaccine hesitancy.
\newblock \emph{{PNAS Nexus}}, 1\penalty0 (4), 2022.
\newblock URL \url{https://doi.org/10.1093/pnasnexus/pgac207}.

\bibitem[Robertson(2018)]{Robertson2018Nov}
R.~Robertson.
\newblock {Partisan Bias Scores for Web Domains}.
\newblock Harvard Dataverse, 2018.
\newblock URL \url{https://doi.org/10.7910/DVN/QAN5VX}.

\bibitem[Robertson et~al.(2023)Robertson, Jiang, Joseph, Friedland, Lazer, and
  Wilson]{Robertson2023}
R.~E. Robertson, S.~Jiang, K.~Joseph, L.~Friedland, D.~Lazer, and C.~Wilson.
\newblock {Auditing Partisan Audience Bias within Google Search}.
\newblock \emph{Proc ACM Hum.-Comput Interact}, 2\penalty0 (CSCW):\penalty0
  1--22, 2023.
\newblock URL \url{https://doi.org/10.1145/3274417}.

\bibitem[Roozenbeek et~al.(2020)Roozenbeek, Schneider, Dryhurst, Kerr, Freeman,
  Recchia, van~der Bles, and van~der Linden]{Roozenbeek2020Oct}
J.~Roozenbeek, C.~R. Schneider, S.~Dryhurst, J.~Kerr, A.~L.~J. Freeman,
  G.~Recchia, A.~M. van~der Bles, and S.~van~der Linden.
\newblock {Susceptibility to misinformation about COVID-19 around the world}.
\newblock \emph{R Soc Open Sci}, 7\penalty0 (10):\penalty0 201199, Oct. 2020.
\newblock URL \url{https://doi.org/10.1098/rsos.201199}.

\bibitem[Rostila(2010)]{Rostila2010BirdsIll}
M.~Rostila.
\newblock {Birds of a feather flock together --- and fall ill? Migrant
  homophily and health in Sweden}.
\newblock \emph{Sociol Health Illn}, 32\penalty0 (3):\penalty0 382--399, Mar.
  2010.
\newblock URL \url{https://doi.org/10.1111/j.1467-9566.2009.01196.x}.

\bibitem[Salathé and Khandelwal(2011)]{Salathe2011Oct}
M.~Salathé and S.~Khandelwal.
\newblock {Assessing Vaccination Sentiments with Online Social Media:
  Implications for Infectious Disease Dynamics and Control}.
\newblock \emph{PLoS Comput Biol}, 7\penalty0 (10):\penalty0 e1002199, Oct.
  2011.
\newblock \doi{https://doi.org/10.1371/journal.pcbi.1002199}.

\bibitem[Shao et~al.(2018)Shao, Ciampaglia, Varol, Yang, Flammini, and
  Menczer]{shao2018spread}
C.~Shao, G.~L. Ciampaglia, O.~Varol, K.-C. Yang, A.~Flammini, and F.~Menczer.
\newblock The spread of low-credibility content by social bots.
\newblock \emph{Nature communications}, 9\penalty0 (1):\penalty0 1--9, 2018.
\newblock URL \url{https://doi.org/10.1038/s41467-018-06930-7}.

\bibitem[Simon and Camargo(2021)]{Simon2021Jul}
F.~M. Simon and C.~Q. Camargo.
\newblock {Autopsy of a metaphor: The origins, use and blind spots of the
  {`}infodemic{'}}.
\newblock \emph{New Media {\&} Society}, 25\penalty0 (8):\penalty0 2219--2240,
  July 2021.
\newblock \doi{10.1177/14614448211031908}.
\newblock URL \url{https://doi.org/10.1177/14614448211031908}.

\bibitem[Smith and Christakis(2008)]{smith2008social}
K.~P. Smith and N.~A. Christakis.
\newblock Social networks and health.
\newblock \emph{Annu. Rev. Sociol}, 34:\penalty0 405--429, 2008.
\newblock URL \url{https://doi.org/10.1146/annurev.soc.34.040507.134601}.

\bibitem[Sontag et~al.(2022)Sontag, Rogers, and Yates]{Sontag2022Mar}
A.~Sontag, T.~Rogers, and C.~A. Yates.
\newblock {Misinformation can prevent the suppression of epidemics}.
\newblock \emph{J R Soc Interface}, 19\penalty0 (188):\penalty0 20210668, Mar.
  2022.
\newblock URL \url{https://doi.org/10.1098/rsif.2021.0668}.

\bibitem[Sooknanan and Comissiong(2020)]{Sooknanan2020Jul}
J.~Sooknanan and D.~M.~G. Comissiong.
\newblock {Trending on Social Media: Integrating Social Media into Infectious
  Disease Dynamics}.
\newblock \emph{Bull Math Biol}, 82\penalty0 (7):\penalty0 1--11, July 2020.
\newblock URL \url{https://doi.org/10.1007/s11538-020-00757-4}.

\bibitem[Tay et~al.(2024)Tay, Lewandowsky, Hurlstone, Kurz, and
  Ecker]{Tay2024Jan}
L.~Q. Tay, S.~Lewandowsky, M.~J. Hurlstone, T.~Kurz, and U.~K.~H. Ecker.
\newblock {Thinking clearly about misinformation}.
\newblock \emph{Commun. Psychol.}, 2\penalty0 (4):\penalty0 1--5, Jan. 2024.
\newblock \doi{10.1038/s44271-023-00054-5}.
\newblock URL \url{https://doi.org/10.1038/s44271-023-00054-5}.

\bibitem[Thaker and Subramanian(2021)]{Thaker2021}
J.~Thaker and A.~Subramanian.
\newblock Exposure to covid-19 vaccine hesitancy is as impactful as vaccine
  misinformation in inducing a decline in vaccination intentions in new
  zealand: Results from pre-post between-groups randomized block experiment.
\newblock \emph{Frontiers in Communication}, 6, 2021.

\bibitem[{United States Census Bureau}(2023)]{UScensusPopClock}
{United States Census Bureau}.
\newblock {Population Clock}, Nov. 2023.
\newblock URL \url{https://www.census.gov/popclock}.
\newblock [Online; accessed 6. Nov. 2023].

\bibitem[Verelst et~al.(2016)Verelst, Willem, and Beutels]{Verelst2016Dec}
F.~Verelst, L.~Willem, and P.~Beutels.
\newblock {Behavioural change models for infectious disease transmission: a
  systematic review (2010-2015)}.
\newblock \emph{J R Soc Interface}, 13\penalty0 (125):\penalty0 20160820, 2016.
\newblock \doi{10.1098/rsif.2016.0820}.
\newblock URL \url{https://doi.org/10.1098/rsif.2016.0820}.

\bibitem[Weng et~al.(2013)Weng, Menczer, and Ahn]{Lilian2013srep}
L.~Weng, F.~Menczer, and Y.-Y. Ahn.
\newblock Virality prediction and community structure in social networks.
\newblock \emph{Sci. Rep.}, 3\penalty0 (2522), 2013.
\newblock \doi{10.1038/srep02522}.
\newblock URL \url{http://dx.doi.org/10.1038/srep02522}.

\bibitem[Yuan et~al.(2023)Yuan, Jahani, Zhao, Ahn, and
  Pentland]{yuan2023implications}
Y.~Yuan, E.~Jahani, S.~Zhao, Y.-Y. Ahn, and A.~S. Pentland.
\newblock {Implications of COVID-19 vaccination heterogeneity in mobility
  networks}.
\newblock \emph{Communications Physics}, 6\penalty0 (1):\penalty0 206, 2023.

\bibitem[Zeng et~al.(2024)Zeng, Chang, and Liu]{Zeng24QMF}
R.~Zeng, X.~Chang, and B.~Liu.
\newblock Evolutionary modeling and analysis of opinion exchange and epidemic
  spread among individuals.
\newblock \emph{Frontiers in Physics}, 12, 2024.

\end{thebibliography}

\newpage
\setcounter{page}{1}
\appendix

\counterwithin{figure}{section}
\counterwithin{table}{section}

\counterwithout{figure}{section}
\counterwithout{table}{section}

\renewcommand{\thefigure}{S\arabic{figure}}
\renewcommand{\thetable}{S\arabic{table}}

\setcounter{figure}{0}

\renewcommand{\thesection}{\arabic{section}}

\renewcommand{\thesubsection}{\thesection.\arabic{subsection}}

\part*{Supplementary information}
\label{sec:si}

\section{Mean-field SMIR model}
\label{sec:meanfield}

For both the ordinary and misinformed subpopulations, the Susceptible Misinformed Infected Recovered (SMIR) model replicates the standard SIR compartments, denoted as $S_O$/$I_O$/$R_O$ and $S_M$/$I_M$/$R_M$, respectively.
SMIR adopts distinct transmission parameters for the misinformed ($\beta_M$) and ordinary ($\beta_O$) groups.
(In the agent-based model, these are proportional to $p_M$ and $p_O$, respectively.) 
The mean-field approximation assumes that the population is well mixed, ignoring the empirical network structure, and that infected individuals from either group ($I_O + I_M$) can potentially infect \textit{anyone} in the susceptible populations.
The mean-field model is governed by the following system of equations:
\begin{equation} \label{eq:system2}
    \left\{
        \begin{array}{@{}l@{}}
            \frac{dS_O}{dt} = -\beta_O S_O (I_O + I_M),
            \; \; 
            \frac{dI_O}{dt} = \beta_O S_O (I_O + I_M) - \gamma I_O,
            \; \; 
            \frac{dR_O}{dt} = \gamma I_O
            \medskip
            \\
            \frac{dS_M}{dt} = -\beta_M S_M (I_O + I_M),
            \; \; 
            \frac{dI_M}{dt} = \beta_M S_M (I_O + I_M) - \gamma I_M,
            \; \; 
            \frac{dR_M}{dt} = \gamma I_M.
        \end{array}
    \right.
\end{equation}
To model homophily, we modify the term $I_O + I_M$ in Eq.~\ref{eq:system2} to account for increased (decreased) contacts within (across) groups, according to the parameter $\alpha \geq 0.5$.
When homophily does not play a role ($\alpha=0.5$), there is an equal probability of interacting with either subpopulation's infected group.
We thus obtain: 
\begin{equation} \label{eq:system3}
    \left\{
        \begin{array}{@{}l@{}}
            \frac{dS_O}{dt} = -2\beta_O S_O (I_O\alpha + I_M(1-\alpha)),
            \; \; 
            \frac{dI_O}{dt} = 2\beta_O S_O (I_O\alpha + I_M(1-\alpha)) - \gamma I_O,
            \; \; 
            \frac{dR_O}{dt} = \gamma I_O
            \medskip
            \\
            \frac{dS_M}{dt} = -2\beta_M S_M (I_O(1-\alpha) + I_M \alpha),
            \; \;  
            \frac{dI_M}{dt} = 2\beta_M S_M (I_O(1-\alpha) + I_M \alpha) - \gamma I_M,
            \; \; 
            \frac{dR_M}{dt} = \gamma I_M.
        \end{array}
    \right.
\end{equation}

Let us denote the proportions of misinformed and ordinary individuals as $\mu$ and $1-\mu$, respectively.
A proportion $\epsilon = 0.001$ of the population is initially infected, split evenly between the ordinary and misinformed groups.
Thus, during the initial state, we have initial values for each compartment: $R_O = R_M = 0$, $S_O = \mu-\frac{\epsilon}{2}$, $S_M = 1-\mu-\frac{\epsilon}{2}$, $I_M = \frac{\epsilon}{2}$, and $I_O = \frac{\epsilon}{2}$.

\begin{figure}[p]
  \centerline{\includegraphics[width=0.8\linewidth]{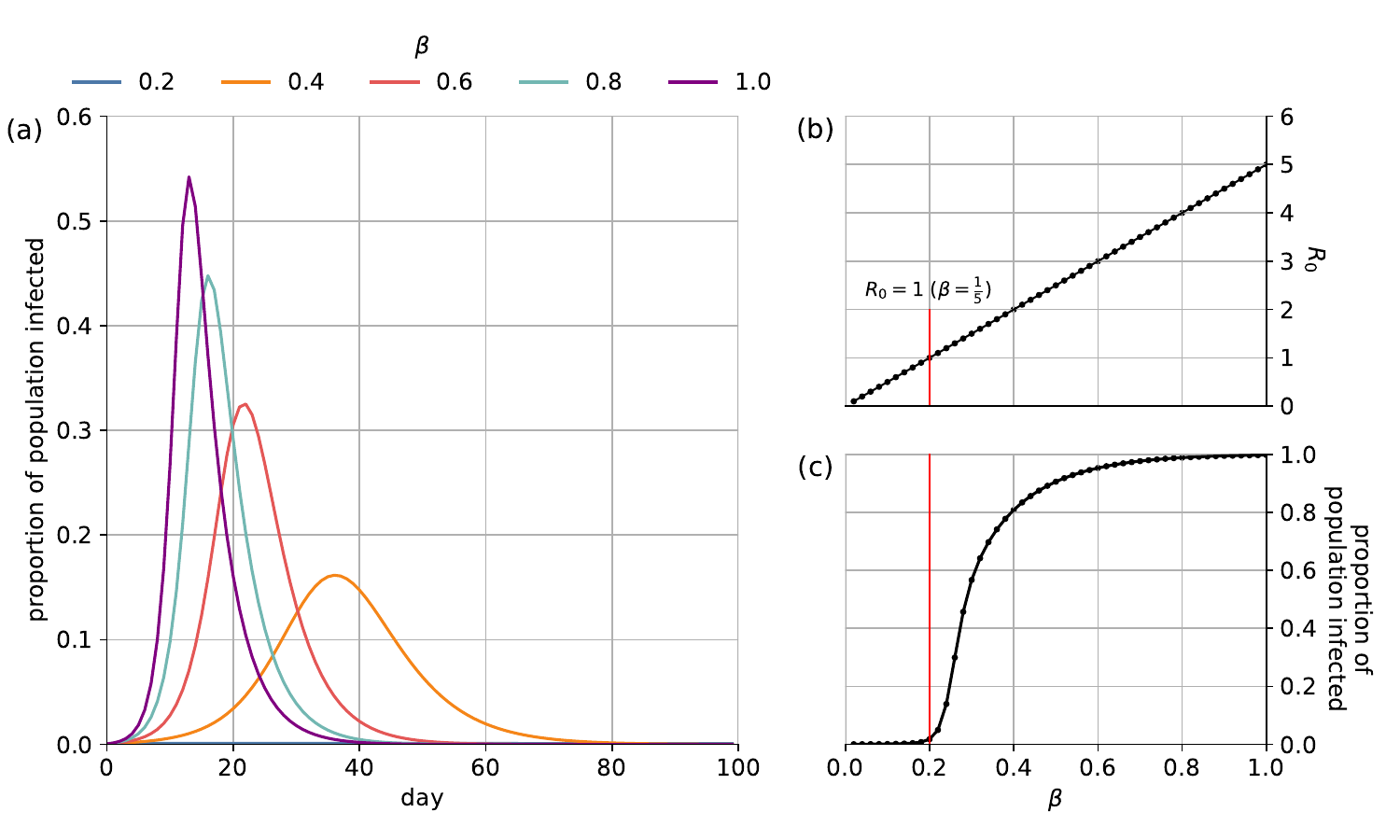}}
  \caption{Reducing the transmission parameter $\beta_O$ decreases the severity of the epidemic.
  We plot (a)~the proportion of the population infected each day, (b)~$R_0$ values for the ordinary population, and (c)~the total proportion of the population infected as $\beta_O$ varies.
  In (a), the curve for $\beta_O = 0.2$ is difficult to see because the proportion of the population infected remains very low throughout the simulation.
  Here we do not consider the role of misinformation or homophily.}
  \label{fig:nohomo_beta}
\end{figure}

To identify a suitable base value for the transmission rate among ordinary susceptibles, we begin by exploring the scenario with no misinformed individuals ($\mu = 1$), setting $\gamma=0.2$ and varying the transmission parameter in the range $0.02 \leq \beta_O \leq 1$.
As is typical of SIR dynamics, Fig.~\ref{fig:nohomo_beta} shows that lower $\beta_O$ values delay and lower the infection peak --- the so-called ``flattening of the curve.''
Lower $\beta_O$ also decreases the total proportion of the population that becomes infected at any point during the epidemic, while higher $\beta_O$ values increase this proportion.
These dynamics are tied to the basic reproduction number $R_0 = \beta / \gamma$: the disease spreading dynamics only reach epidemic levels when $R_0 > 1$, such that an infected individual infects more than one other person on average. This happens when $\beta_O > 0.2$.
As $R_0$ increases, the infection spreads more quickly, the peak infection day occurs sooner, and the proportion of the population that is ultimately infected increases.

\begin{figure}[p]
  \centerline{\includegraphics[width=0.8\linewidth]{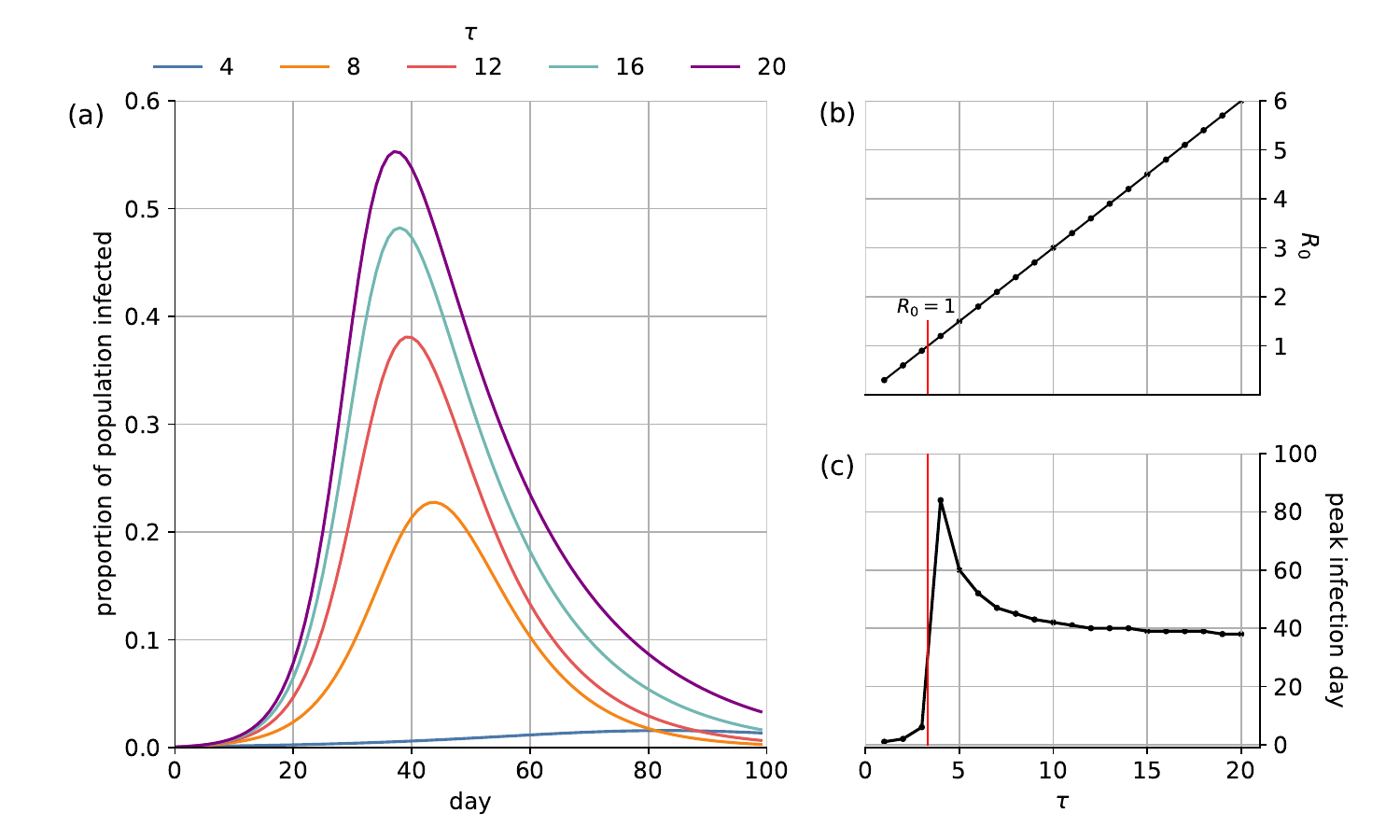}}
  \caption{ 
  Effects of varying the recovery rate. 
  We plot (a)~the proportion of the population infected each day, (b)~$R_0$ values for the ordinary population, and (c)~the total proportion of the population infected as a function of the number of days to recover, $\tau$.
  Here we do not consider the role of misinformation or homophily.}
  \label{fig:nohomo_recovery}
\end{figure}

We now explore the effect of the recovery rate, again in the scenario with no misinformation or homophily, by setting $\beta_O = 0.3$ and varying the recovery period $\tau=1/\gamma$ between 1 and 20 days. 
Fig.~\ref{fig:nohomo_recovery} shows that when $\tau < 4$, $R_0 < 1$ and the disease does not reach epidemic proportions.
At this level, the epidemic takes a long time to reach its peak ($\approx$80 days).
Increasing $\tau$ means that individuals remain infected longer, so the population gets infected faster and the peak infection is reached more rapidly.

In summary, the effects of varying the transmission and recovery parameters are predictable: lower $\beta$ and higher $\gamma$ ``flatten the curve'' and reduce the negative outcomes of an infection. 
Based on these explorations we set $\tau = 5$ ($\gamma=0.2$) to align with quarantine recommendations from the CDC\cite{cdc_isolation_2023}. 
We further set $\beta_O = 0.3$ such that the basic reproduction number is $R_0 \ge \beta_O / \gamma = 1.5$ to ensure epidemic spread within the ordinary population.

\section{Mean-field analyses}

\begin{figure}[t]
  \centerline{\includegraphics[width=\linewidth]{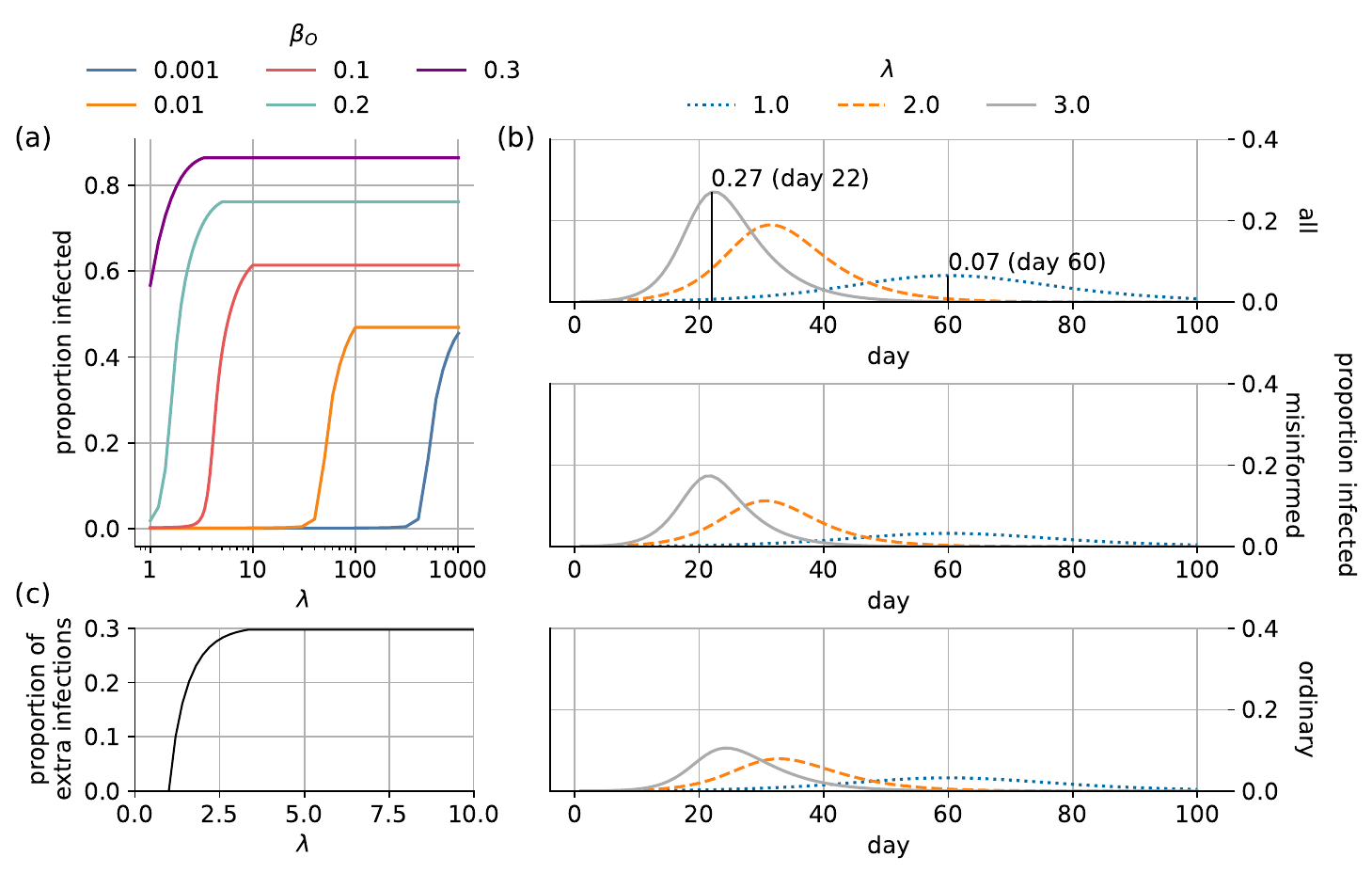}}
  \caption{ 
  Increasing $\lambda = \beta_M / \beta_O$ accelerates and amplifies the infection. We use  $\gamma = 0.2$, and $\mu=0.5$.
  (a)~Overall proportion of the population infected as a function of $\lambda$, for different values of $\beta_O$.  
  (b)~Proportion of the population infected on each day, for different values of $\lambda$ using $\beta_O = 0.3$. 
  (c)~Extra proportion of the total population that is infected as a function of $\lambda$ ($\beta_O = 0.3$).
  }
  \label{fig:nohomo_lambda}
\end{figure}

To explore the effects of risky behaviors by misinformed individuals, let us assume two equally-sized subpopulations ($\mu = 1/2$) and introduce the scaling factor $\lambda = \beta_M / \beta_O \ge 1$. 
Fig.~\ref{fig:nohomo_lambda}(a) illustrates the increasing negative impact of the misinformed subpopulation on the disease-spreading dynamics as $\lambda$ becomes larger. 
If the ordinary population has very low $\beta_O$, $\lambda$ has to be very high for the misinformed population to have an effect.
On the other hand, if $\beta_O$ is large enough, increasing $\lambda$ leads to a ceiling effect, as $\beta_M$ cannot exceed one.
The social cost associated with the more risky behaviors by the misinformed group is passed on to the whole network. 
For example, when $\lambda = 3.0$, peak infection for the entire population is reached 38 days earlier than in the $\lambda = 1$ case (day 22 vs. 60; Fig.~\ref{fig:nohomo_lambda}(b)), leading to an additional 29.3\% of the population becoming infected (Fig.~\ref{fig:nohomo_lambda}(c)).

\begin{figure}[t]
  \centerline{\includegraphics[width=\linewidth]{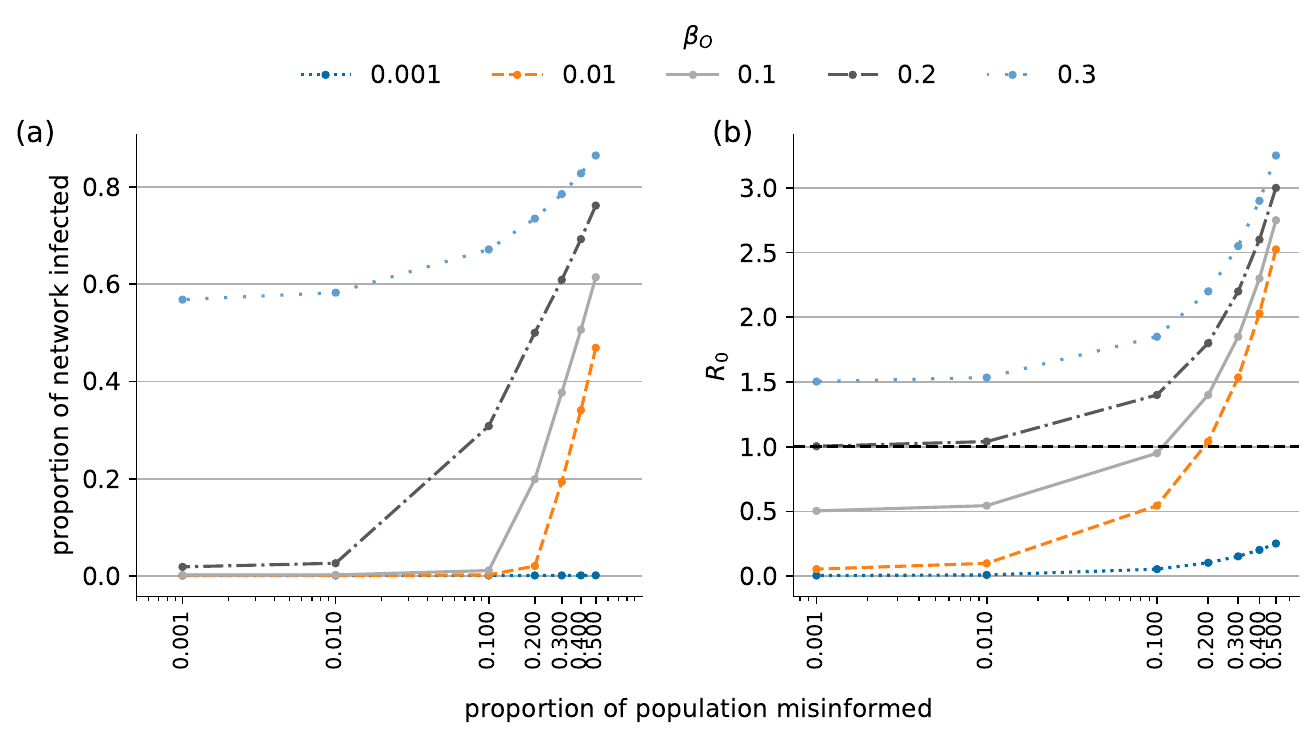}}
  \caption{ 
  Increasing the initial proportion $\mu$ of the population that is misinformed, as well as $\beta_O$, increases (a)~the size of the network that becomes infected and (b)~the average $R_0$ across the population. 
  Here, we fix $\lambda=100$ to match the ratio used in the main text.
  }
  \label{fig:misinfo_prop}
\end{figure}

We further explore how the initial size $\mu$ of the misinformed population affects the total proportion of the network that ultimately gets infected. We consider various values of $\beta_M$ and $\beta_O$ such as to capture the same $\lambda = \beta_M / \beta_O = p_M / p_O = 100$ as in the main text. 
When $\mu$ and $\beta_O$ are both low, the misinformed population has no impact on the infection (Fig.~\ref{fig:misinfo_prop}(a)), as $R_0<1$ (Fig.~\ref{fig:misinfo_prop}(b)). 
However, increasing either parameter crosses the epidemic threshold ($R_0>1$) so that a significant portion of the population gets infected. 

\section{Effect of homophily}

Let us explore the effect of homophily among the ordinary and misinformed subpopulation networks.
Homophily means that infected individuals are more likely to interact with (and infect) susceptibles from the same subpopulation (ordinary or misinformed) than the other group.
The degree of homophily is modeled by a parameter $\alpha$. When $\alpha=0.5$, individuals are equally likely to interact within and across groups (no homophily), whereas $\alpha = 1$ is the case when homophily is strongest and the subpopulations do not interact with each other (see Methods for details).

\begin{figure}
  \centerline{
  }
  \centerline{\includegraphics[width=\linewidth]{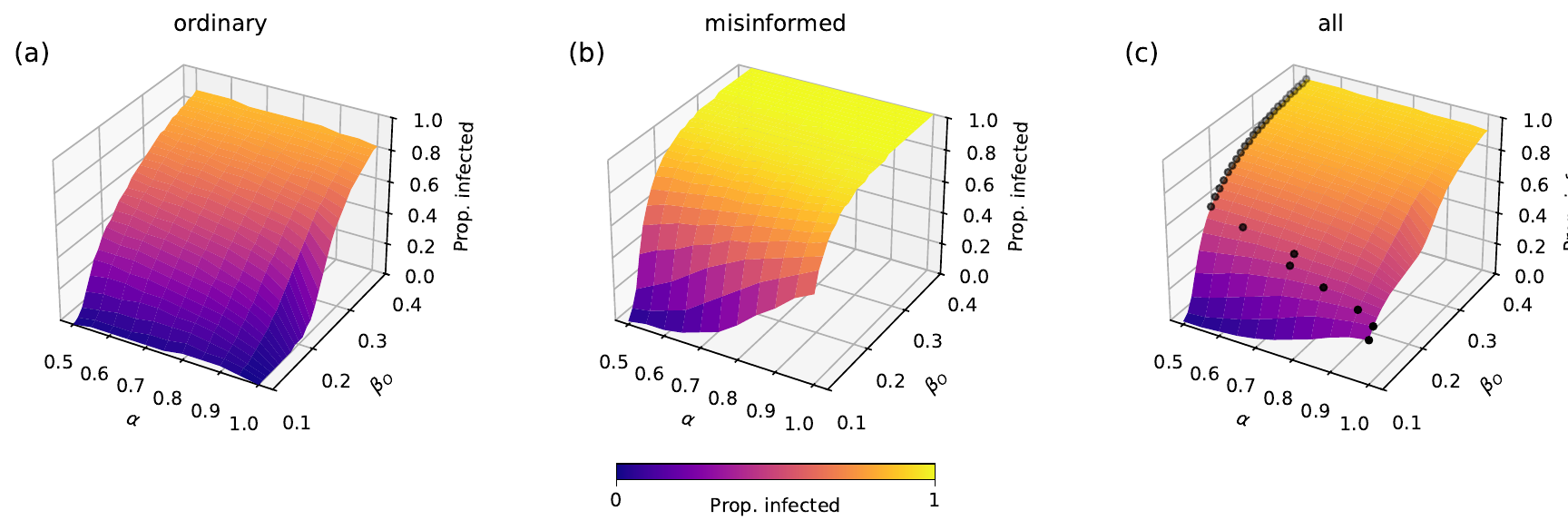}}
  \caption{ 
  Homophily in the contact network worsens the infection among misinformed individuals, especially for lower transmission rates. 
  The combined effects of transmission and homophily parameters,  $\beta_O$ and $\alpha$, are examined with the mean-field approximation when $\lambda = 3$, $\gamma = 0.2$, and $\mu=0.5$. We plot the proportions of infected individuals in (a) the ordinary population, (b) the misinformed population, and (c) the overall population. The maximum proportion of the overall population infected for each $\beta_O$ is marked with a black dot. When the transmission rate is sufficiently high, homophily benefits the entire population but harms the misinformed group.}
  \label{fig:homo_alpha}
\end{figure}

Fig.~\ref{fig:homo_alpha} illustrates the effects of homophily ($0.5 \le \alpha \le 1$) for different levels of ordinary transmission ($0.1 \le \beta_O \le 0.4$). 
For less infectious disease (low $\beta_O$), increasing homophily significantly harms the misinformed group (Fig.~\ref{fig:homo_alpha}(b)): the infection remains confined within this group. There is no discernible effect on the ordinary population as long as the two groups interact; when they do not ($\alpha = 1$), we observe a sharp reduction in infections (Fig.~\ref{fig:homo_alpha}(a)). 
For $0.12 < \beta_O < 0.16$, peak infection scenarios coincide with intermediate homophily levels, as indicated by the black dots in Fig.~\ref{fig:homo_alpha}(c) \cite{PhysRevLett.113.088701}. 
Under these conditions, while increased homophily decreases infections in the general populace, it significantly worsens outcomes for the misinformed group (compare Fig.~\ref{fig:homo_alpha}B and C). 
As $\beta_O$ increases further, while nearly the entire misinformed population becomes infected regardless of $\alpha$ (Fig.~\ref{fig:homo_alpha}(b)), high homophily shields the full population (Fig.~\ref{fig:homo_alpha}(c)): ordinary individuals have a lower risk of becoming infected through interactions with misinformed individuals. 
In summary, homophily offers greater protection to the ordinary group by isolating misinformed communities, which suffer a greater disease burden, exacerbating health disparities \cite{Rostila2010BirdsIll, smith2008social}.

\section{Robustness analyses}

\begin{figure}
    \centering
    \includegraphics[width=0.6\linewidth]{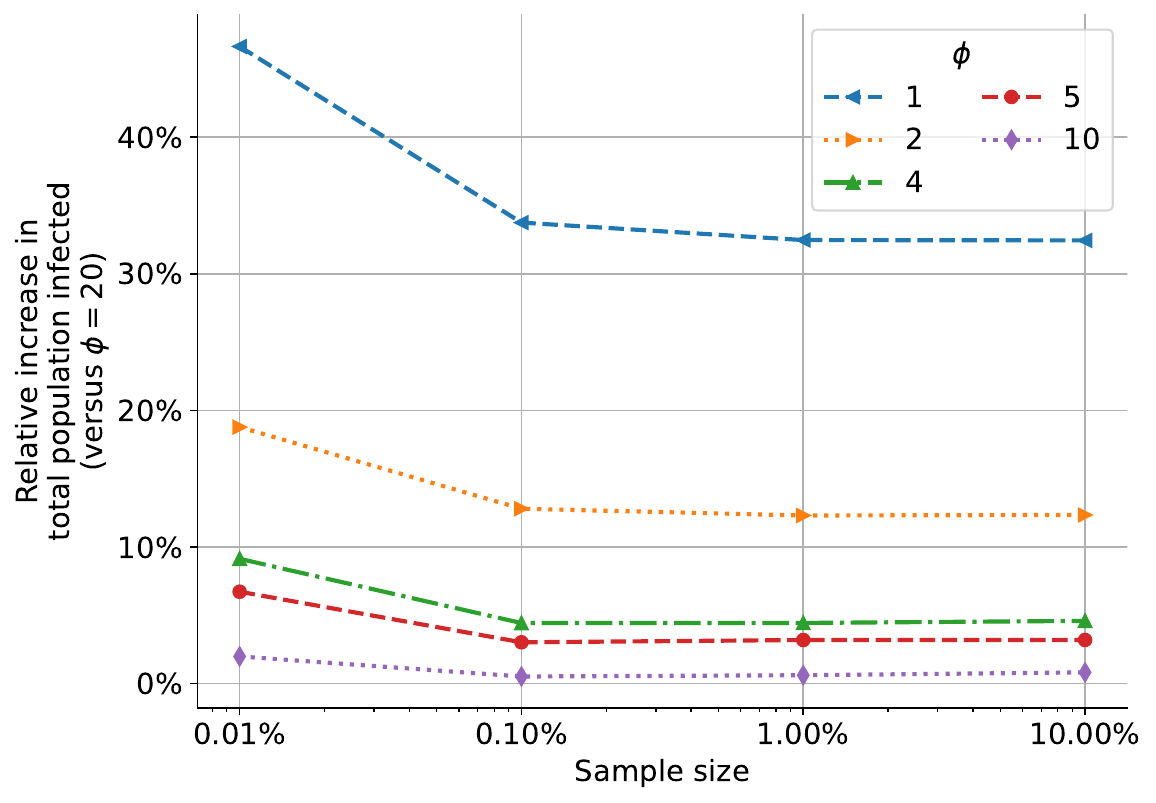}
    \caption{Relative increase in the mean total population infected as a function of the sampling size utilized in the contact network creation process. The $\phi = 20$ scenario, in which the fewest nodes in the network are misinformed, is utilized as the baseline.}
    \label{fig:relative_increase}
\end{figure}

To test the robustness of the main results (Fig.~\ref{fig:effect_of_phi}C) with respect to the sample size used to construct the contact network, all simulations were rerun after generating contact networks based on the different sampling percentages between 0.01\% and 10\%. 
Fig.~\ref{fig:relative_increase} shows the relative increase in the percentage of the population that becomes infected as a function of the linear threshold $\phi$, using the best-case scenario in which the fewest nodes in the network are misinformed as the baseline.
We observe a substantial decrease in the effect of misinformation as the sampling size grows to 1\%. 
However, sample sizes above 1\% return nearly identical results. 
We conclude that using a sample size of 10\% (as reported in the main text) is sufficient to rule out any size-induced bias. 

\begin{figure}
    \centering
    \includegraphics[width=0.8\linewidth]{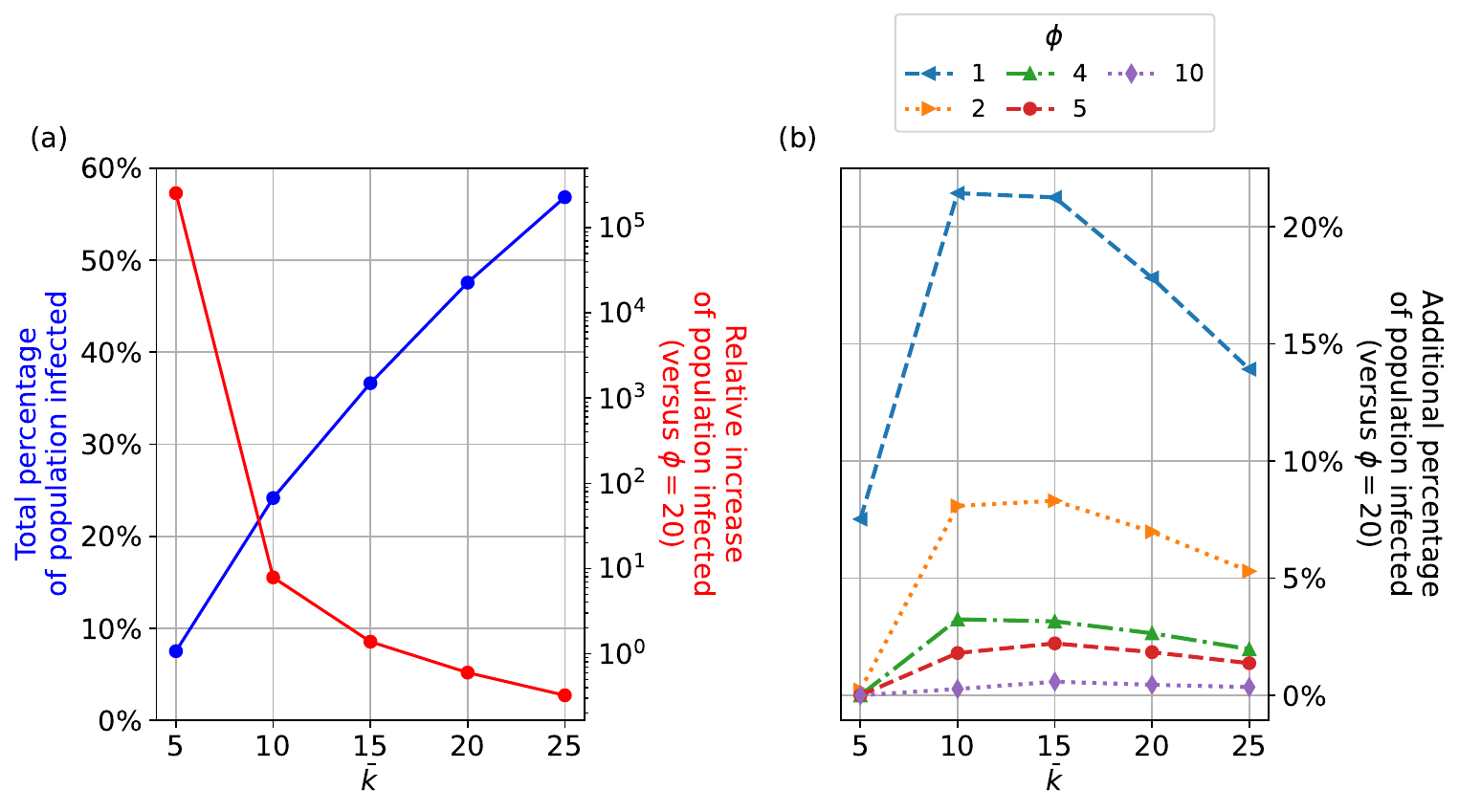}
    \caption{ 
    Effects of average contact network degree $\bar{k}$ on infection dynamics. (a) Infected individuals ($\phi = 1$) as a percentage of the overall population and relative to the baseline condition $\phi = 20$, in which the fewest nodes in the network are misinformed. (b) Additional percentages of infected population relative to the baseline condition $\phi = 20$.}
    \label{fig:avg_degree}
\end{figure}

Fig.~\ref{fig:avg_degree} illustrates the impact of the contact network density ($\bar{k}$) on infection dynamics, for $5 \le \bar{k} \le 25$.
We consider this range because $\bar{k}=25$ represents pre-pandemic daily social contacts while $\bar{k}=5$ represents COVID-19 lockdown conditions\cite{LiuSocialContactPatterns}. 
As expected, Fig.~\ref{fig:avg_degree}(a) demonstrates that higher $\bar{k}$ leads to increased infections through the population, since the higher contact density provides more opportunities for transmission. 
But while a larger percentage of the overall population is infected, the relative effect of misinformed individuals decreases. 
This is because, at higher $\bar{k}$ values, the infected population is already substantial even in the low-misinformation ($\phi=20$) baseline.
The combined effect of these two opposing trends, as shown in Fig.~\ref{fig:avg_degree}(b), is that the additional percentage of infected individuals relative to the $\phi=20$ scenario reaches a maximum for some intermediate $\bar{k}$.
Fig.~\ref{fig:avg_degree}(b) also shows that, consistent with our primary findings, increasing $\phi$ (misinformed resilience) decreases the infected population. 
In our main analysis we focus on $\bar{k}=25$ and model an effective reduction of contacts by decreasing the $p$ parameter. 

\end{document}